\documentclass[aps,pre,twocolumn,superscriptaddress]{revtex4-2}
\usepackage{amsmath} 
\DeclareMathOperator{\sgn}{sgn}
\usepackage{graphicx}
\usepackage{hyperref}% add hypertext capabilities
\usepackage{xcolor} %colorful text
\graphicspath{ {./Figures/} }
\usepackage{siunitx}

\begin{document}

%\preprint{}

\title{Vibrational Transportation of Deformable Axisymmetric Particles}

\author{Marina E. Terzi}
\email[Contact author: ]{marina.e.terzi@gmail.com}
\homepage[ORCID: ]{https://orcid.org/0000-0001-9399-4806}
\affiliation{Laboratoire d'Acoustique de l'Université du Mans, Av. Olivier Messiaen, 72085, Le Mans, France}

\author{Vladislav V. Aleshin}%
\affiliation{Institut d'Electronique, de Microélectronique et de Nanotechnologie, UMR CNRS 8520, Université de Lille, Av. Poincaré, CS 60069, 59652, Villeneuve d'Ascq, France}

\author{Jules Ghesquière}%
\affiliation{Institut d'Electronique, de Microélectronique et de Nanotechnologie, UMR CNRS 8520, Université de Lille, Av. Poincaré, CS 60069, 59652, Villeneuve d'Ascq, France}

\author{Vincent Tournat}%
\affiliation{Laboratoire d'Acoustique de l'Université du Mans, Av. Olivier Messiaen, 72085, Le Mans, France}

\date{\today}

\begin{abstract}
A particle on a substrate supporting a surface acoustic wave can experience horizontal drift excited by the dry friction force. The effect is referred to as vibrational transportation, or as a surface acoustic wave motor, and is used in a number of industrial applications. A traditional theory of vibrational transportation considers a particle as a material point moving on a rigid substrate. A more realistic representation is a contact model based on Cattaneo-Mindlin (also called Hertz-Mindlin) mechanics applicable to an axisymmetric deformable particle. The contact zone in this case is not a point, but rather a circle that generally contains a smaller circle of stick and a surrounding annulus of slip. A recent semi-analytical extension of the Cattaneo-Mindlin solution called the Method of Memory Diagrams (MMD) allows one to compute the hysteretic friction force for an arbitrary loading history in terms of contact displacements, and, subsequently, to numerically solve the equations of motion.
Depending on the materials' and excitation parameters, the particle can stay in permanent contact with the substrate or experience multiple jumps. In the former case, the particle can drift in a horizontal direction due to asymmetric sliding condition created by oscillating normal and tangential contact forces. In other words, during each wave period, the particle advances and recedes with different efficiencies, which finally results in a drift. The drift can occur in the wave propagation direction and against it, and require a specific choice of system's parameters.
In the regime of multiple jumps, directed horizontal motion is also possible. However, it is governed by a completely different mechanism based on synchronization between the wave period and rebounding events. There exist cases where the rebound occurs once per period and consistently at the same phase. During each rebound, the particle gets horizontal momentum of the same sign (against the wave propagation direction). The value of this momentum depends on the horizontal velocity mismatch between the particle and the substrate; therefore, at the beginning of the process, the particle moves with an acceleration that decreases and finally disappears. Exactly the same type of motion against the wave has been observed in our preliminary experiments. In other cases, the time of flight of the particle and the wave period are uncorrelated, thus resulting in a chaotic motion. 
We also demonstrate that a point mass in the same situation behaves differently. In particular, in a regime of permanent contact, negative and positive sliding are equilibrated, which produces no drift. In addition, multiple rebounds of a point mass are always chaotic, at least for fully conservative collisions.
In conclusion, the deformable particle model can be a better guide for various applications, such as particle micropositioning or acoustic dust cleaning. 
\end{abstract}

\maketitle


\section{Introduction\label{sec:intro}}

It is well known that small objects (particles) posed on vibrating surfaces can experience directed tangential motion~\textemdash~a phenomenon often referred to as \textit{vibrational transportation}. This effect is of interest for a variety of applications, such as shaking chutes, conveyors, and vibratory feeders. These devices are used in the food and pharmaceutical industries for delivering powders and grains, as well as for packaging, sorting, and precise medical dosing~\cite{erdesz_experimental_1988,mracek_system_2005}. Potential applications of the principle include mobile mechanisms for surgery~\cite{kim_forward_2020}, devices for sensor cleaning and sample sorting, e.g. in exploratory space missions~\cite{hui_vibrational_2024,bao_transport_2009}, vibration-based dust cleaning systems for plate-like structures such as solar panels~\cite{abd-elhady_new_2024}, size-based particle sorting~\cite{dunst_vibration-assisted_2018}, programmable patterning~\cite{kopitca_programmable_2021}. The requirements for vibrational systems can range from precise, selective manipulation or micropositioning to rapid, large-scale particle displacement (e.g. for cleaning applications). Since the 1980s, attempts have been made to estimate the transport rate in such devices~\cite{hongler_periodic_1989}. 

The effect of horizontal particle motion on vibrating surfaces has been known for centuries. In particular, standing waves excited in vibrating plates were observed to move sand grains towards the nodes of these waves (Chladni plates~\cite{chladni_discoveries_1787,zhou_controlling_2016}). Recently, a reverse Chladni effect has been observed, where particles move toward the anti-nodes. This behavior is attributed to the currents generated by the vibrating plate in the surrounding fluid~\cite{latifi_motion_2019} or air~\cite{van_gerner_air-induced_2011}. By adjusting the parameters of the standing waves, one can shift the nodes and anti-nodes, causing the particles to follow them. A striking example of this phenomenon is a particle escaping a maze~\cite{latifi_motion_2019}. Particles subjected to standing waves can also form complex patterns such as letters or other symbols~\cite{zhou_controlling_2016}.

Generally, a particle drift in a specific direction requires the asymmetry in the excited particle-substrate contact system. An exhaustive list of possible asymmetries, along with an extensive set of analytical solutions, is given in a fundamental book on vibrational mechanics~\cite{blekhman_basic_1994}. The considered asymmetry cases include anisotropic friction coefficient, internal particle motion, inclined surfaces, traveling surface acoustic waves (SAWs), etc.

The situation in which particles are moved by the friction force induced by traveling acoustic waves localized at interfaces~\textemdash such as the Rayleigh or Lamb modes~\textemdash deserves additional attention. In such waves, vertical and horizontal displacements are phase-shifted, creating the asymmetry required for directed motion. Horizontal particle motion is driven by a friction force arising from acoustic displacements. Examples of practical applications of the effect include Rayleigh ultrasonic motors~\cite{kurosawa_friction_1996}, SAW devices that manipulate powders~\cite{bao_transport_2009,sherrit_mechanism_2010,takizawa_manipulation_2021,saiki_transporting_2021} and droplets~\cite{bao_transport_2009,brunet_droplet_2010,bar-cohen_high-speed_2012}. 

To describe particle motion under the action of acoustically generated friction force, it is necessary to define a contact model. Three approaches are possible: a point mass on a deformable substrate, a point mass on a rigid substrate that, however, supports an acoustic wave, and a deformable particle of finite size on a deformable substrate. The first case assumes the use of the Boussinesq solution~\cite{boussinesq_application_1885,landau_theory_1986} for a concentrated force applied to an elastic half-space. The solution for the normal displacement field exhibits a singularity at the contact point and is therefore unusable for solving the particle dynamics problem. To the best of our knowledge, in a fully deformable contact system, only vertical motion has been considered so far~\cite{ranganath_nayak_contact_1972,perret-liaudet_resonance_1998,perret-liaudet_response_2006}. Thus, most existing theoretical approaches to vibrational transport focus on the material point model. 

Depending on the problem parameters, the particle can either stay in permanent contact with the substrate or undergo multiple jumps (hopping motion). Material point dynamics has been extensively explored in~\cite{blekhman_vibrational_1964} for both permanent contact and hopping cases, considering a specific kind of vibrations and for various vertical and horizontal restitution coefficients. Motion induced by a traveling Rayleigh wave has been studied by~\cite{verma_particle_2013} using a similar approach. 

As for the hopping material point, its normal motion has been analyzed by several researchers in terms of stability. In particular, the derivation in~\cite{luck_bouncing_1993} showed that, depending on the situation, the vertical motion of a point mass jumping on the vibrating surface can be stable or chaotic. A typical example of stable motion is periodic hopping, where rebounds occur every $n$ wave periods at the same phase. In contrast, in the case of dynamic chaos, rebounds are not synchronized with the wave period and therefore can occur at any phase. 
Introducing tangential motion leads to particle drift in the stable case, as tangential contact interactions produce horizontal velocity increments of the same sign, resulting in coherent velocity accumulation. However, when the normal motion is chaotic, the tangential motion is also chaotic, since the velocity increments are uncorrelated.

Another peculiar type of point mass' vertical motion is chattering, i.e.~multiple rebounds with rapidly decaying amplitude. When the rebounds are not perfectly elastic, it is possible to observe even complete chattering or locking, where an infinite number of impacts with decreasing amplitude occur within a finite time. Sometimes chattering alternating with flights of $n$ wave periods duration leads to a stable trajectory~\cite{luck_bouncing_1993}. Some periodic trajectories with $n=1$ may transition to chaos~\cite{luo_dynamics_1996}.

A curious case of tangential drift of a point mass is observed on a vertically oscillating wavy platform having a sinusoidal profile~\cite{halev_bouncing_2018}. In this regime called ``walking motion'', the horizontal speed of the point mass equals the profile wavelength divided by the platform oscillation period. Partially inelastic collisions with the platform are required for the effect to emerge.

The authors of ~\cite{ragulskis_transport_2008} consider a situation where a point mass falls on a platform in which a Rayleigh wave is excited. Depending on systems' parameters such as restitution coefficients characterizing rebounds, viscosity of air, particle's mass, wave frequency and amplitude, as well as the initial coordinate and velocity of the particle, a variety of dynamic regimes are observed. These regimes include chaotic and regular jumps, as well as chattering finally transforming into permanent contact. It is also shown that chaos can appear by the transition from periodic jumping through period doubling.

Accounting for the deformability of the contact system introduces new effects. Analytical results for a vertical motion of a Hertzian particle~\cite{hertz_uber_1881,landau_theory_1986} on an oscillating platform revealed the existence of a contact resonance~\cite{ranganath_nayak_contact_1972}. At frequencies much lower than the resonance one, the rigid particle approximation remains valid. However, near resonance and at higher frequencies, the dynamics of a deformable sphere differ significantly. In addition, a second-order subharmonic resonance exists for the Hertzian particle~\cite{perret-liaudet_response_2006}. In this case, excitation at twice the resonant frequency is converted to the fundamental frequency through a specific nonlinear process. A second-order superharmonic resonance, typical of nonlinear systems, has also been reported~\cite{perret-liaudet_resonance_1998}. 

A numerical model~\cite{morita_simulation_1999} for the horizontal motion of a normally deformable sphere on a substrate supported by a Rayleigh wave suggested that motion against the wave is possible for a specific system configuration (geometry, materials, wave frequency and amplitude). The model also explained the observed accelerated motion with eventual velocity saturation~\footnote[1]{However, this model disregards transitions to the stick state and contains a sign error in the expression for sliding friction force, which calls its results into question.}. 

In this paper, we intend to take a step forward by considering the normal and tangential motion of a deformable particle. The tangential behavior of the contact system is described using the Method of Memory Diagrams (MMD, see~\cite{aleshin_method_2015,delrue_two_2018}) a recent generalization of the Cattaneo-Mindlin (also called the Hertz-Mindlin) solution~\cite{cattaneo_sul_1938,mindlin_elastic_1953}. This solution links contact forces and displacements for the contact of two spheres under constant normal compression followed by tangential loading. The presence of friction in the form of the Coulomb friction law results in a partial slip~\textemdash an annular region of slip propagating inward from the contact boundary, while a circular interior remains stuck (see also~\cite{jager_axi-symmetric_1995}). The MMD~\cite{aleshin_method_2015} extends this solution to construct a semi-analytical formalism applicable to excitation by an oblique shift with an arbitrary time dependence. In addition to partial slip, the MMD extension~\cite{delrue_two_2018} also describes total sliding, as well as contact loss (particle detachment). The model is quasi-statical, neglecting the inertia of the strained material near the contact and related impact effects.

Other representatives of a semi-analytical family of methods in contact mechanics include the method of dimensionality reduction ~\cite{popov_method_2015,popov_handbook_2019}. The approach is based on a mathematical mapping of the contact problem onto contact between a rigid body and a set of equally spaced springs. Applied to a variety of cases, the approach has also been used for an approximate description of the impact of an elastic sphere with an elastic half-space~\cite{willert_impact_2016}.

For both the deformable particle and material point models, the normal component of the particle-substrate collision is assumed to be conservative to limit the number of system parameters. The tangential component of contact interaction is lossy and hysteretic due to friction.

Similarly to the point-mass model, the deformable contact system exhibits both continuous contact and bouncing modes. However, the account for deformability significantly alters the particle's behavior. In the continuous contact mode, the deformable particle can drift in the direction of the wave and against it. The mechanism in this case can be called asymmetric sliding and consists in the alternation of advancing and receding stages having different sliding paths during each wave period. At the same time, we found for the point mass that these paths in the opposite directions equilibrate, therefore no drift occurs. In the bouncing (hopping) mode, the point mass experiences a chaotic behavior when all subsequent flight times and, consequently, wave phases during collisions are different. As a result, the tangential momentums received by the particle due to the frictional interaction at the moments of collisions are not correlated. For the deformable particle, this chaotic situation also occurs; however, an autostabilizing regime may emerge when the times of flight match the wave period. Phased tangential momentums produce drift that is demonstrated to be against the wave. Experimental observations of drift against the wave direction~\cite{morita_simulation_1999, bao_transport_2009,takizawa_manipulation_2021,behera_design_2019} indicate that the incorporation of deformability is essential to accurately describe vibrational transport. 

This paper is organized as follows. The problem of a particle on a substrate excited by a traveling Rayleigh wave is formulated in Section~\ref{sec:equations}, followed by the description of the contact model in Section~\ref{sec:MMD}. In Section~\ref{sec:results}, possible motion modes are presented via phase diagrams, similar to~\cite{blekhman_vibrational_1964, verma_particle_2013}, together with typical trajectory examples. Section~\ref{sec:conclusions} deals with various drift mechanisms and comparison of the deformable particle and material point models. Some practical recommendations for vibrational transport and dust-cleaning systems are also given. Our earlier results were reported in~\cite{aleshin_contact_2025,terzi_hopping_2025}.

\section{Governing equations for a deformable axisymmetric particle on a substrate excited by the Rayleigh wave\label{sec:equations}}

\begin{figure}[htbp]
\centering   
\includegraphics[width=0.45\textwidth]{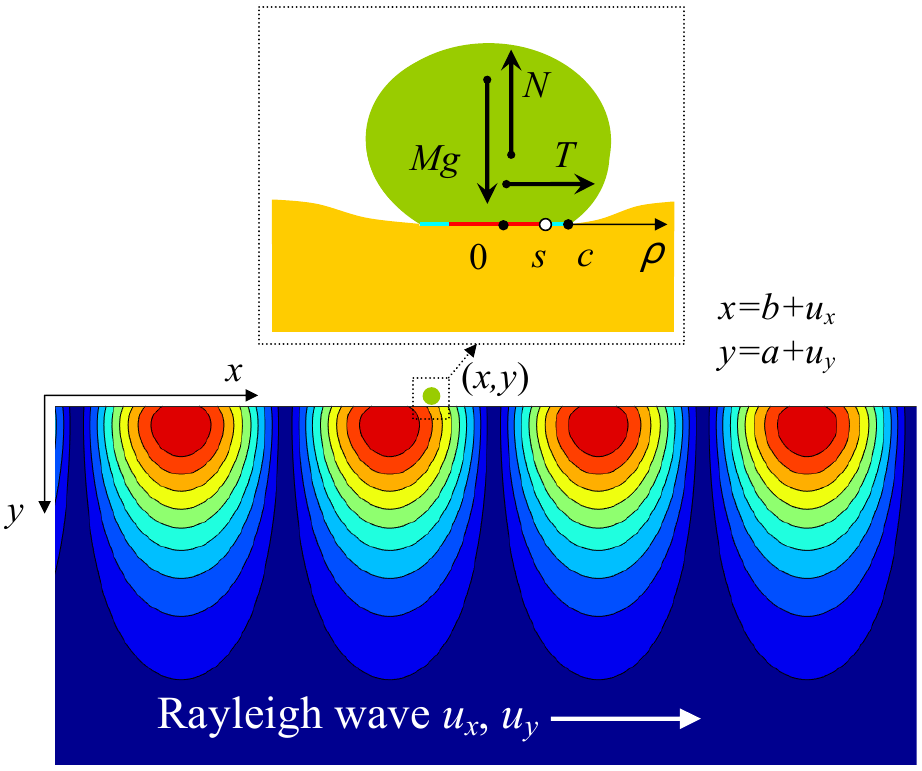}%{Fig1_geom-1.eps}
    \caption{Geometry of a deformable particle moving on a substrate in which a Rayleigh wave is excited. $x$ and $y$ are the coordinates of the particle in the laboratory reference frame, $b$ and $a$ are its displacements relative to the moving substrate.\label{fig:problem_formulation}} 	
\end{figure}

Here we consider a deformable particle posed on a deformable substrate in which a traveling harmonic Rayleigh wave is excited. Unlike a material point, this particle has a finite size and can undergo various types of motion. However, the Cattaneo-Mindlin model on which our approach is based accounts only for translational degrees of freedom, excluding rolling, tilting, and torsion. Following these restrictions, we write equations of motion for the coordinates $x(t)$ and $y(t)$ assuming that the other degrees of freedom are either not excited or constrained by the chosen experimental geometry (e.g., using three agglomerated particles or a truncated sphere). Denoting the normal and tangential components of the contact force as $N$ and $T$, respectively, the equations of motion take the form (see Fig.~\ref{fig:problem_formulation})

\begin{subequations} \label{eq:motion}
\begin{align}
m\ddot x &= -T\textrm{ ,} \label{eq:motion_x}\\
m\ddot y &= Mg-N\label{eq:motion_y}\textrm{ .}
\end{align}
\end{subequations}

Here the term $Mg$ accounts for gravity and a possible external pre-compression force $P$ which may, for instance, be of electrostatic nature:
\begin{equation}
Mg=mg+P\textrm{ ,}
\end{equation}

where $m$ is the actual mass of the particle and $M$ is a parameter with the dimension of the mass. The contact force $T$ is equal in magnitude and opposite in direction to the friction force, following the classical convention used by~\cite{mindlin_elastic_1953}.

The contact force components ($N$, $T$) induce contact displacements ($a$, $b$) defined relative to an initial unstrained state. In the presence of an acoustic wave with displacement components ($u_x$, $u_y$), the particle's coordinates ($x$, $y$) in the laboratory frame of reference are given by

\begin{subequations} \label{eq:xbya}
\begin{eqnarray}
x &= u_{x}+b\textrm{ ,} \label{eq:xb}\\
y &= u_{y}+a\label{eq:ya}\textrm{ .}
\end{eqnarray}
\end{subequations}

Here, $a$ and $b$ can be interpreted as coordinates of the particle in the reference frame attached to the substrate which itself moves according to the prescribed wave displacements ($u_x$, $u_y$) relative to the laboratory reference frame. The displacements in a Rayleigh wave~\cite{viktorov_rayleigh_1966} are determined as

\begin{widetext}
\begin{subequations} \label{eq:rayleigh}
\begin{align}
u_x(t)&=-A_yr(\nu)\sin(2\pi f t-k_Rx)\approx -A_yr(\nu)\sin(2\pi f t)\textrm{ ,} \label{eq:rayleigh_x}\\
u_y(t)&=A_y\cos(2\pi f t-k_Rx)\approx A_y\cos(2\pi f t)\label{eq:rayleigh_y}\textrm{ .}
\end{align}
\end{subequations}
\end{widetext}

where the ratio of horizontal to vertical ($A_y$) amplitudes is given by
\begin{equation}
r(\nu)=\dfrac{k_R\left(1-\dfrac{2q_Rs_R}{k_R^2+s_R^2}\right)}{q_R\left(1-\dfrac{2k_R^2}{k_R^2+s_R^2}\right)}
\end{equation}

with 
\begin{equation}
 \left\{\begin{array}{@{}l@{}@{}}
q_R^2=k_R^2-k_l^2\textrm{, }
s_R^2=k_R^2-k_t^2\textrm{, }\\
k_l=\frac{2\pi}{c_l}\textrm{, }
k_t=\frac{2\pi}{c_t}\textrm{ .}
\end{array}\right.
\end{equation}

Here, $c_l$ and $c_t$ are longitudinal and transverse wave velocities in the substrate material, respectively.

In Eq.~\eqref{eq:rayleigh}, $k_Rx$ is neglected since, as we shall see in Section~\ref{sec:horizontal}, the typical distance traveled by the particle during one wave period is much smaller than the Rayleigh wavelength.

Equations of motion~\eqref{eq:motion} must be supplemented with initial conditions. A natural choice for these would be to start from the rest state, given by:

\begin{subequations} \label{eq:initial}
\begin{align}
&x=\dot x=0\textrm{ ,} \label{eq:initial_x}\\
&y=a_0 \textrm{, } \dot y=0\label{eq:initial_y}\textrm{ .}
\end{align}
\end{subequations}

where $a_0$ is the normal displacement created by the static force $Mg$ in the absence of the wave. 

However, this definition contradicts the stationary form of Eq.~\eqref{eq:rayleigh} for the Rayleigh wave that assumes no rest state. To mitigate this controversy, we use a ramp function:

\begin{equation}
           \Gamma(t)=1-e^{-\frac{t^2}{\tau^2}}
\end{equation}

to turn the wave on progressively from $0$ to a certain amplitude. The ramp function appears as a factor in Eq.~\eqref{eq:xbya}, so that the final form of the Rayleigh wave displacement becomes

\begin{subequations}
  \begin{align}
    u_x(t) &= \Gamma(t) A_yr(\nu) \sin(2\pi f t) \nonumber\textrm{,} \\
    u_y(t) &= \Gamma(t) A_y \cos(2\pi f t) \nonumber\textrm{.}
  \end{align}
\end{subequations}

The activation time of the ramp function $\tau$ should be much longer than the wave period ($\tau\gg 2\pi/f$) to minimize the non-physical distortion $e^{-\frac{t^2}{\tau^2}}$ in the Rayleigh wave expression.

To solve the equations of motion, one has to specify the contact forces as functions of the contact displacements according to the selected model.

\section{Force-displacement relationships for an axisymmetric particle\label{sec:MMD}}

The force-displacement relationship consists of normal and tangential components. The normal one is supposed to be fully reversible and independent of the tangential one:
\begin{equation}
N=N(a).
\end{equation}
For a sphere compressed against a half-space $N(a)$ is given by the classical Hertz solution. For non-spherical but axisymmetric profiles, the solution is provided in~\cite{jager_axi-symmetric_1995}.

The tangential motion of an axisymmetric particle is governed by the Cattaneo-Mindlin (also known as Hertz-Mindlin) dynamics. The classical solution~\cite{cattaneo_sul_1938,mindlin_elastic_1953} applies when the contact system is first loaded in the normal direction with a force $N$ and then subjected to a tangential force $T$. Normal loading creates a circular contact zone of radius $c$. After applying $T$, a slip annulus forms in the region $s < \rho < c$, while the central region $\rho < s$ remains stuck. If $T$ increases from 0 to $\mu N$ (where $\mu$ is the friction coefficient), the slip zone appears and expands inward from the periphery until it reaches the center ($\rho = 0$), resulting in total sliding. 

An elegant form of the Cattaneo-Mindlin solution~\cite{jager_axi-symmetric_1995} can be written using the notation $N_c$ and $a_c$ which stands for the normal force and displacement required to form a contact zone of radius $c$:

\begin{subequations} \label{eq:Tb_nc_ac}
\begin{align}
T&=T_c^s=\mu(N_c-N_s)\textrm{ ,}\label{eq:Tnc} \\
 b&=b_c^s=\theta\mu(a_c-a_s) \textrm{.}\label{eq:bac}
\end{align}
\end{subequations}

Here $N_s$ and $a_s$ are the force and displacement required to create contact with the radius $s$, and $\theta$ is the ratio of tangential to normal contact stiffness~\cite{popov_method_2015}:

\begin{equation}
\theta = \dfrac{\dfrac{(2-\nu)(1+\nu)}{2E}+\dfrac{(2-\nu_p)(1+\nu_p)}{2E_p}}{\dfrac{1-\nu^2}{E}+\dfrac{1-\nu_p^2}{E_p}}\,.
\label{eq:theta}
\end{equation}

In the above expression, $E$ and $\nu$ are the Young modulus and the Poisson ratio of the substrate material, respectively, while $E_p$ and $\nu_p$ refer to those of the particle material.

The Method of Memory Diagrams (MMD) is a recent generalization of the above approach, applicable when the loading history is more complex (e.g., corresponds to an acoustic excitation). The drive parameters of the model are the displacements $a$ and $b$, which can evolve arbitrarily over time. The algorithm operates with complex combinations of $T^s_c$ and $b^s_c$, incorporating various coefficients $s$ and $c$, which are stored in memory or erased according to the prescribed rules. Depending on the loading history and the current displacement values $b$ and $a$, four cases can occur:
\begin{enumerate}
 \item Partial slip when $s<c$, with $T$ determined by the MMD algorithm~\cite{aleshin_method_2015};
 \item Full stick when $s=c$ , with $T$ determined by the MMD algorithm~\cite{aleshin_method_2015};
 \item Total sliding when $s=0$, with $T=\sgn(\dot{b})\mu N$;
 \item Contact loss when $c=0$, with $T=0$.
\end{enumerate}

The tangential force $T$, calculated according to the above rules, is denoted here as

\begin{equation}
T = T_{\textrm{MMD}}(b,a)\,.
\end{equation}

The above expression includes two material constants: the friction coefficient $\mu$ and the stiffness ratio $\theta$ (see Eq.~\eqref{eq:theta}).

It is convenient to rewrite the equations of motion Eq.~\eqref{eq:motion} in the normalized form:

\begin{subequations} \label{eq:motion_norm}
\begin{align}
m^*\ddot{x^*} &= -\mu T^*\textrm{ ,} \label{eq:motion_norm_x}\\
 m^*\ddot {y^*} &= 1-N^*(a^*)\textrm{ ,}  \label{eq:motion_norm_y}
\end{align}
\end{subequations}

and Eq.~\eqref{eq:xbya} as:

\begin{subequations} \label{eq:xbya_norm}
\begin{align}
\theta\mu b^*&=x^*-R^*(t^*)A_y^*r(\nu)\sin 2\pi t^*\textrm{ ,} \label{eq:xb_norm}\\
a^*&=y^*-R^*(t^*)A_y^*\cos 2\pi t^*\textrm{ ,}  \label{eq:ya_norm}
\end{align}
\end{subequations}

in which

\begin{equation}
 \left\{\begin{array}{@{}l@{}}
N^*=\dfrac{N}{N_0}\textrm{ , }T^*=\dfrac{T}{\mu N_0}\textrm{ , }m^*=\dfrac{a_0mf^2}{Mg}\textrm{ , }t^*=ft\\
(a^*,x^*,y^*,A_y^*)=\dfrac{(a,x,y,A_y)}{a_0}\textrm{ , }b^*=\dfrac{b}{a_0\theta\mu}\textrm{ , }
  \end{array}\right.\
\label{eq:normalization}
\end{equation}

and $N_0=N(a_0)$. Here, the normalized ramp function $\Gamma^*(t^*)\equiv \Gamma(t)$.

The initial conditions Eq.~\eqref{eq:initial} transform into

\begin{subequations} \label{eq:initial_norm}
\begin{align}
 x^*=\dot x^*=0\textrm{ ,} \label{eq:initial_norm_x}\\
 y^*=1\textrm{ , }\dot y^*=0\textrm{.}\label{eq:initial_norm_y}
\end{align}
\end{subequations}

An important advantage of the normalized equations~\eqref{eq:motion_norm},~\eqref{eq:xbya_norm},~\eqref{eq:initial_norm} is that they contain only two principal parameters: $m^*$, representing the effective mass, and $A_y^*$, representing the wave amplitude. The other parameters, $\theta$, $\mu$, and $\nu$, can be considered as material constants. In this work, we do not distinguish between static and dynamic friction coefficients.

In practice, typical Poisson’s ratios range from $\nu_{\textrm{min}} = 0.1$ to $\nu_{\textrm{max}} = 0.4$, which corresponds to a relatively narrow band of $\theta$-values between 1.0555 and 1.3333. The numerical results of this paper are obtained for a realistic value of $\theta$ = 1.2; slight variations in $\theta$ do not considerably affect the results. At the same time, the friction coefficient $\mu$ varies over a broader range, from 0.01 to 1, in a wide variety of practical cases~\cite{serway_physics_1996}.

Eqs.~\eqref{eq:motion_norm} and~\eqref{eq:xbya_norm} with initial conditions Eq.~\eqref{eq:initial_norm} have been numerically solved using the $4^{th}$-order Adams-Bashforth method~\cite{hairer_solving_2000}. This method performs explicit time stepping from known values defined on four previous time steps, in contrast to common algorithms of Runge-Kutta type, which use only one previous time step but require knowledge of intermediate values defined between time grid nodes. Examples of various motion regimes for specific values of the essential parameters $m^*$ and $A_y^*$ are given below. However, before presenting these results, the normal contact problem should first be solved.

\section{Results\label{sec:results}}
\subsection{Vertical motion of a Hertzian particle}

The normal problem for two elastic axisymmetric bodies is solved in~\cite{galin_contact_1961,jager_axi-symmetric_1995}. The Hertz solution for the normal force-displacement relationship between two spheres in contact is a particular case and is given by~\cite{popov_method_2015}
\begin{equation}
   N(a)=\begin{cases}
    \frac{4}{3}E^*R^{*1/2}\left(\frac{a}{2}\right)^{3/2}, & \text{if $a>0$}\\
    0, & \text{if $a \leq 0$} \textrm{ .}
  \end{cases}
\end{equation}
with the effective modulus $E^*$ and radius $R^*$ defined as

\begin{equation}
 \left\{\begin{array}{@{}l@{}}
\frac{2}{E^*}=\frac{1-\nu^2}{E}+\frac{1-\nu_p^2}{E_p}\textrm{ ,}\\
\frac{2}{R^*}=\frac{1}{R}+\frac{1}{R_p}\textrm{ .}
\end{array}\right.
\label{eq:ERstar}
\end{equation}

Here, $E_p$ and $\nu_p$ are Young’s modulus and Poisson’s ratio of the particle material, and $R_p$ denotes the radius of the particle. The second body is a half-space with radius $R=\infty$, and the Hertz solution becomes

\begin{equation}
  N(a)=\begin{cases}
    \frac{2}{3}E^*R^{*1/2}\left(\frac{a}{2}\right)^{3/2}, & \text{if $a>0$}.\\
    0, & \text{if $a \leq 0$}.
  \end{cases}
  \label{eq:Na_sphere_plane}
\end{equation}

The normalized version of Eq.~\eqref{eq:motion_norm_y} 
\begin{equation}
a^*=y^*-\Gamma^*(t^*)A_y^*\cos (2\pi t^*)
\label{eq:astar}
\end{equation}
describes vertical motion excited by the normalized normal force

\begin{equation}
  N^*(a^*)=\begin{cases}
   a^{*3/2}, & \text{if $a^*>0$}.\\
    0, & \text{if $a^* \leq 0$}
  \end{cases}
  \label{eq:Na_norm}
\end{equation}

with the initial conditions Eq.~\eqref{eq:initial_norm_y}. Note that the contact system determined by Eqs.~\eqref{eq:motion_norm_y},~\eqref{eq:astar},~\eqref{eq:Na_norm},~\eqref{eq:initial_norm_y} is perfectly elastic (lossless).

\begin{figure*}[htbp]
	\includegraphics[width=\textwidth]{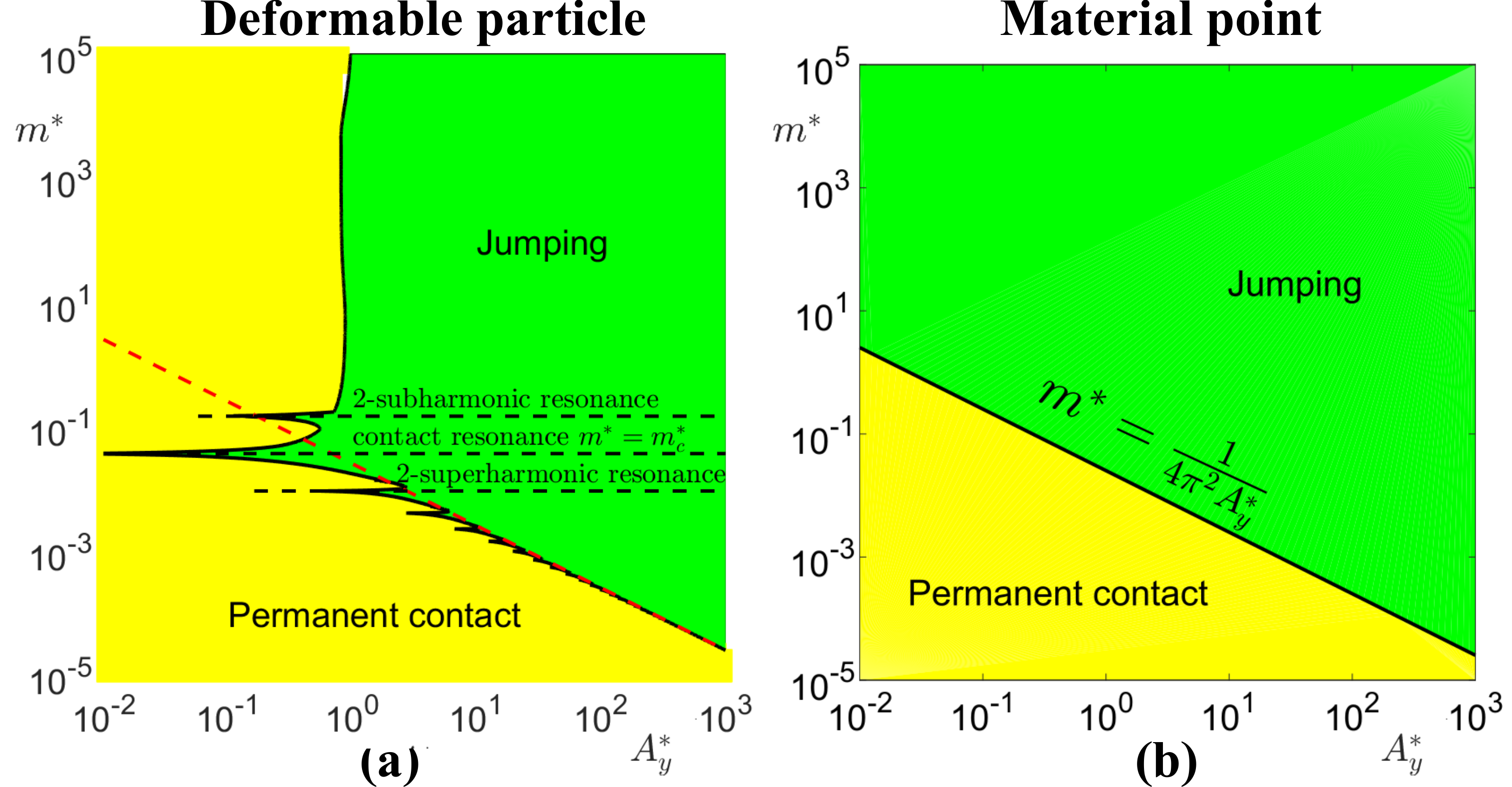}%{Figures/Fig2_vertical_motion.eps}
		\caption{Phase diagrams for (a) deformable particle (Hertzian sphere) and (b) material point with domains corresponding to two modes of vertical motion: jumping and permanent contact. The boundary between the domains (black curve) is obtained numerically in (a) and analytically in (b), following~\cite{verma_particle_2013}. Fundamental, subharmonic and superharmonic resonances are indicated by horizontal dashed lines in (a); in (b) there are no resonances. The inclined dashed line in (a) corresponds to the black line in (b) and is included for comparison.  \label{fig:vertical}}	
\end{figure*}

Equations~\eqref{eq:motion_norm_y},~\eqref{eq:astar},~\eqref{eq:Na_norm},~\eqref{eq:initial_norm_y} can easily be solved numerically. There are two possibilities for a particle governed by these equations: to remain in contact with the substrate throughout the observation time or to detach from it when $a^*$ becomes negative. Obviously, if a detachment occurs at some amplitude $A_y^*$, it will also occur for any higher amplitude. Therefore, for any normalized mass $m^*$, there exists a unique threshold amplitude $A_{y}^{*(J)}$ that separates the jumping and permanent contact regimes in the $({A_y^*,m^*})$ functional space. The delimiting curve $A_{y}^*=A_y^{*(J)}(m^*)$ has been obtained numerically using the 4$^{th}$ order Runge-Kutta method for a deformable sphere (see Fig.~\ref{fig:vertical}(a)).

The curve $A_y^*=A_{y}^{*(J)}(m^*)$ exhibits several minima corresponding to the resonances of the Hertzian sphere: the fundamental contact resonance~\cite{ranganath_nayak_contact_1972}, the 2$^{nd}$ order subharmonic resonance~\cite{perret-liaudet_resonance_1998}, the 2$^{nd}$ order superharmonic resonance~\cite{perret-liaudet_response_2006}, and higher-order superharmonic resonances. The fundamental resonance occurs at $m^*=m_c^*$ that approximately equals~\num{3.69d-2}, where $A_{y}^{*(J)}(m^*)$ reaches a minimum of approximately~\num{5.4d-3}. This implies that even at a very low wave amplitude, the particle starts jumping.
This resonance phenomenon can be explained as follows. In a steady state $(\Gamma^*(t^*)\equiv1)$, the equation of motion Eq.~\eqref{eq:motion_norm_y} can be linearized by assuming that $N^*(a^*)$ oscillates around its mean value of 1 which is determined by gravity and pre-compression. Introducing

\begin{equation}
\xi = y^*-1\textrm{, }|\xi|\ll1 
\end{equation}
yields the basis for this linearization.

The equation of motion then takes the form
 \begin{equation}
m^*\ddot{\xi} = 1-(1+\xi-A_y^*\cos(2\pi t^*)^{3/2}\textrm{.}
\label{eq:motion_xi}
\end{equation}	

Further, by assuming that the wave amplitude is also small, $|A_y^*\ll1|$, Eq.~\eqref{eq:motion_xi} can be linearized by keeping only first-order terms:
\begin{equation}
\ddot{\xi}+\omega_0^{*2}\xi=\frac{3A_y^*}{2m^*}\cos(2\pi t^*)\textrm{,}
\label{eq:motion_xi_lin}
\end{equation}
where the natural frequency of free oscillations equals
 \begin{equation}
 \omega_0^*=\sqrt{\frac{3}{2m^*}}\,.
 \label{eq:Hertz_frequency}
 \end{equation}

 The harmonic oscillator Eq.~\eqref{eq:motion_xi_lin} is in resonance when the natural frequency is tuned to the external source frequency $2\pi$, which finally leads to $m_{c}^*=\frac{3}{8\pi^2}\approx$~\num{3.8d-2}\textemdash a value close to the numerically obtained result, $m_c^*\approx$~\num{3.69d-2}. 

A remarkable feature of Hetzian particle motion is the subharmonic resonance. Using the method of multiple scales (see~\cite{perret-liaudet_response_2006}), it was shown that exciting the Hertzian contact at twice the fundamental frequency $\omega_0$ (given in Eq.~\eqref{eq:Hertz_frequency}) leads to steady-state oscillations at $\omega_0$.

Other peaks of the delimiting curve in Fig.~\ref{fig:vertical}(a) correspond to higher-order superharmonic resonances at frequencies $n\omega_0$. This multiple resonance behavior arises from the elastic force being a nonlinear function of displacement (with a power $3/2$ for the Hertzian contact; other cases are considered in~\cite{nayfeh_response_1983}). More analytical results on the normal motion of harmonically excited Hertzian contacts can be found in~\cite{ranganath_nayak_contact_1972,hess_normal_1991,perret-liaudet_resonance_1998,perret-liaudet_response_2006}.

For a material point on an undeformable substrate, the analogous delimiting curve can be found analytically~\cite{verma_particle_2013}. The zones of permanent contact and of jumping/hopping are depicted in Fig.~\ref{fig:vertical}(b), separated by a straight line on a double logarithmic scale that exhibits no resonances. A notable property of the rigid contact system is that even at an infinitely small amplitude, there exists a frequency range in which the particle detaches. In other words, in the presence of high-frequency noise, the point mass cannot remain at rest and jumps all the time. In contrast, a deformable particle cannot jump at an infinitely small amplitude (see~Fig.~\ref{fig:vertical}(a)).

Nayak showed that for a Hertzian sphere, when $m^*> m_c^*$, the deformability of the contact system significantly affects its behavior, whereas for $m^*\ll m_c^*$, the deformability is negligible~\cite{ranganath_nayak_contact_1972}. This conclusion is confirmed by the delimiting curves in Fig.~\ref{fig:vertical}(a) and (b), which nearly coincide at low $m^*$ but differ considerably at high $m^*$.

In the next section, we explore possible horizontal motion types for a deformable particle both in the permanent contact (Section~\ref{sec:contact_def}, cases 1-3) and in hopping modes (Section~\ref{sec:jump_def}, cases 4-8), and compare them to the behaviors of a material point (Section~\ref{sec:mp}, cases 9-10). 

\subsection{Horizontal motion modes\label{sec:horizontal}}

\begin{figure*}[htbp]
\includegraphics[width=\textwidth]{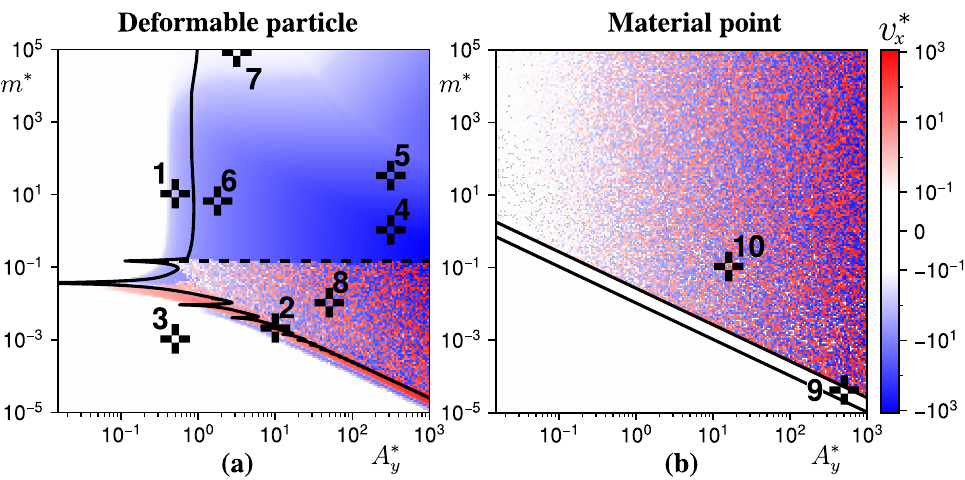}%{Figures/Fig3_Phase_DiagramsV2.eps}
\caption{Dimensionless tangential velocity (see Eq.~\eqref{eq:colormap}) of a deformable Hertz-Mindlin sphere (a) and a material point (b) in the functional space $(A_y^*,m^*)$, averaged over 2000 wave periods. Red and blue colors represent motion in the direction of wave propagation and against it, respectively. Points with $|v_x^*|<v^*_0$ (Eq.~\eqref{eq:colormap}) are shown in white. The solid line separates permanent contact motion from hopping. Here, $\mu=0.3$. Numbers 1\textendash10 indicate different motion types discussed below.  \label{fig:velocity_MMD}} 
\end{figure*}

A synthetic result of our entire study is presented in Fig.~\ref{fig:velocity_MMD}. Without detailing the trajectories and mechanisms of horizontal motion (discussed in depth below), this figure illustrates its efficiency, defined as the total distance covered divided by the observation time. In other words, Fig.~\ref{fig:velocity_MMD} shows the average dimensionless horizontal velocity, $ \langle v_x^* \rangle$ with $v_x^* = \frac{\textrm{d}x^*}{\textrm{d}t}$, plotted over the functional space $(A_y^*, m^*)$ for the deformable contact system (a) and the rigid one (b). 

Across different values of $A_y^*$ and $m^*$, the average horizontal velocity varies in a wide range. To efficiently represent this range, the logarithmic scale is used. Specifically, the color-coded quantity $C$ shown in Fig.~\ref{fig:velocity_MMD} is defined as follows:

\begin{equation}
C=\begin{cases}
   \displaystyle\frac{\log_{10}{(v_x^*/v_0^*)}}{\displaystyle\max\limits_{v_x^*>0}  \Bigg[\displaystyle\log_{10}{(v_x^*/v_0^*)}\Bigg]}\textrm{,} & \text{if $v_x^*>v_0^*$ ,}\\
    0\textrm{,} & \text{if $-v_0^*\le v_x^*\le v_0^*$ ,}\\
     \displaystyle\frac{-\log_{10}{(-v_x^*/v_0^*)}}{\max\limits_{v_x^*<0} \Bigg[\displaystyle\log_{10}{(-v_x^*/v_0^*)}\Bigg]}\textrm{,} & \text{if $v_x^*<-v_0^*$ .}
  \end{cases}
  \label{eq:colormap}
\end{equation}
Positive and negative velocities are plotted in red and blue, respectively. The parameter $v_0^*$ determines the contrast in Fig.~\ref{fig:velocity_MMD}. A higher value of $v_0^*$ highlights only the most efficient drifts, filtering out slower motions. In contrast, a lower $v_0^*$ reveals a broader range of velocities, including moderate and weak drift. Here, we use a relatively low value $v_0^*=0.1$, to capture nearly all drift cases, even those with low efficiency.

It is now appropriate to justify the omission of the term $k_R x$ in Eq.\eqref{eq:rayleigh} introduced in Section~\ref{sec:equations}. This can be done by comparing the highest drift velocity $|v_x^*|\sim~$\num{e3} at point 4 ($m^*=1$, $A_y^*=10^{2.5}$) in Fig.~\ref{fig:velocity_MMD} with the Rayleigh wave speed. Using Eqs.~\eqref{eq:normalization}, the dimensional drift velocity for a deformable particle is given by:

\begin{equation}
v_x = a_0 f v_x^*\textrm{ .}
\end{equation}

Assuming no pre-compression ($m=M$) and substituting typical values for a spherical steel particle (density $\rho_p=8000$~$\textrm{kg}/{\textrm{m}^3}$, $E_p=200$~GPa, $\nu_p=0.3$, $R_p=10^{-3}$~\unit{\metre}) on a glass substrate (density $\rho=2500$~$\textrm{kg}/{\textrm{m}^3}$, $E=70$~GPa, $\nu=0.3$), we obtain $a_0\approx~$\num{9d-10}~\unit{\metre} and $f\approx 10^5$~\unit{\Hz} that corresponds to a physical drift velocity $v_x\approx 9$~$\textrm{cm}/\textrm{s}$. This is about \num{5d4} times lower than the Rayleigh wave speed in glass with the above parameters, $c_R\approx$~4600~\unit{\metre/\second}. This example shows that even a rapidly drifting particle does not experience a significant wave phase shift during its motion. 

The functional diagrams in Fig.~\ref{fig:velocity_MMD}(a) and (b) reveal a number of remarkable features. First, in the permanent contact regime, the deformable particle can drift in both directions (points 1 and 2), while the point mass experiences no drift (point 9). Second, in the jumping mode, a deformable particle can drift against the direction of wave propagation (points 4, 5, and 6), or fall into a ``motley'' zone (point 8) of chaotic motion, where small variations in $A_y^*$ or $m^*$ lead to changes in drift direction. These motley regions indicate sensitivity to initial conditions and are directly related to the stability of motion. In contrast, for a point mass, no consistent drift is observed in the jumping regime; the entire jumping motion zone appears motley. Finally, large portions of the permanent contact zone for the deformable particle are white, indicating negligible or absent horizontal motion.

In the following, various motion modalities are examined in greater detail using examples labeled 1-10 in the diagrams in Fig.~\ref{fig:velocity_MMD}.

\subsubsection{Deformable particle in permanent contact\label{sec:contact_def}}
\paragraph*{\textbf{1) Drift against the wave.}}

\begin{figure*}[htbp]
	\includegraphics[width=\textwidth]{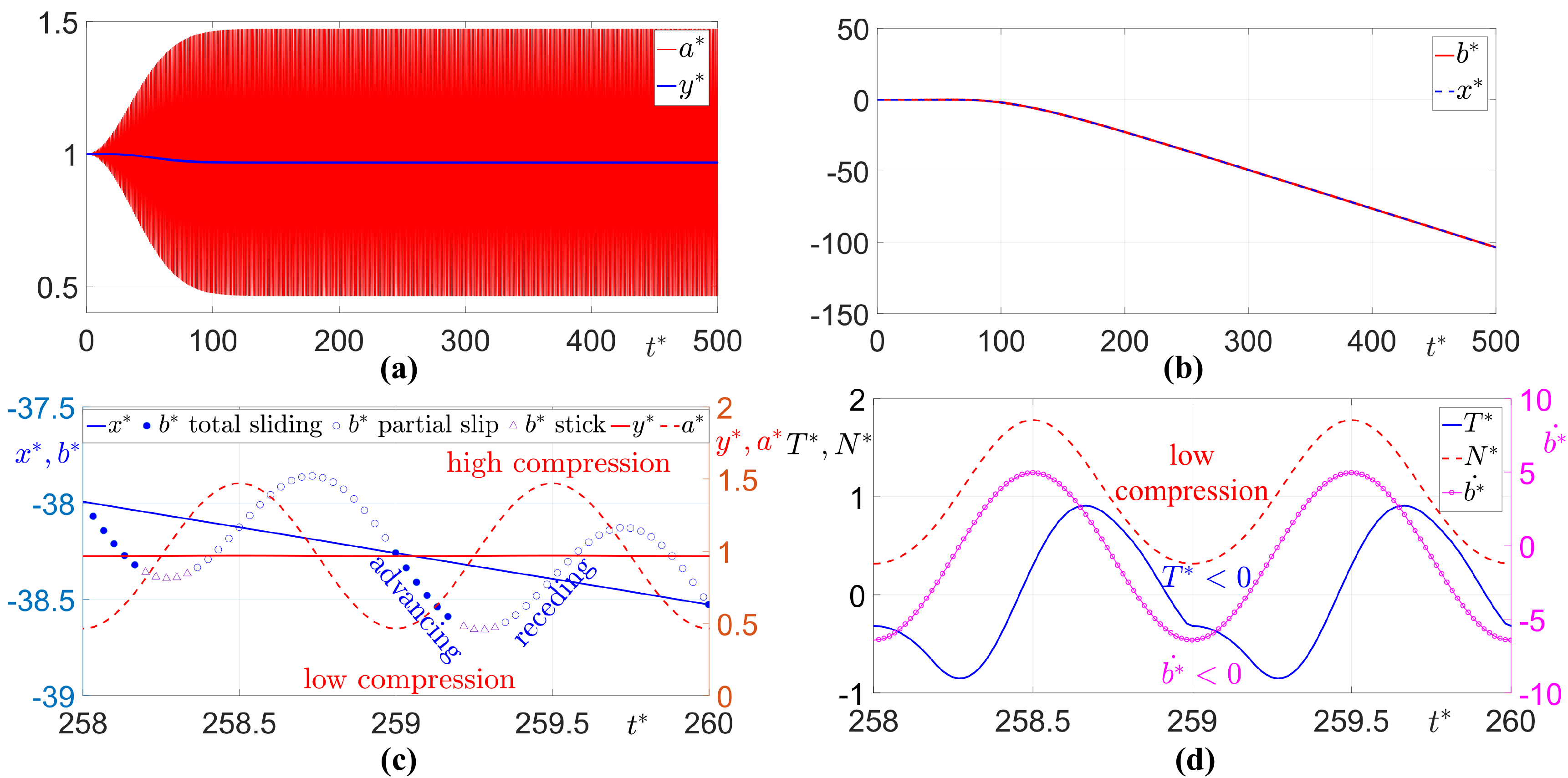}%{Figures/Fig4_against_lgAy-0.3_lgm1.eps}
		\caption{Drift against the direction of wave propagation for a deformable particle in permanent contact (example for $A_y^*=10^{-0.3}$, $m^*=10^{1}$; point 1 in Fig.~\ref{fig:velocity_MMD}). (a) Normal displacements $a^*(t^*)$ and $y^*(t^*)$. (b) Tangential displacements $b^*(t^*)$ and $x^*(t^*)$. (c) Fragments of the above curves. The advancing stage corresponds to total sliding at low compression, while during the receding stage the particle remains stuck at least partially. (d) Normal and tangential forces $N^*(t^*)$, $T^*(t^*)$, and horizontal velocity $\dot{b^*}(t^*)$ relative to the substrate. During total sliding, both $T^*$ and $\dot{b^*}$ are negative, resulting in negative drift direction.  \label{fig:against}}	
\end{figure*}

Motion of this type corresponds to a vertical blue band in the functional diagram Fig.~\ref{fig:velocity_MMD}(a), located to the left of the black delimiting curve (point 1), where $m^*>m^*_c$. The corresponding particle dynamics is illustrated in Fig.~\ref{fig:against}(a)\textendash(d).

Panel (a) in Fig.~\ref{fig:against} illustrates the vertical motion. When the particle mass exceeds the resonance value $m^*_c$, the particle deformability effect becomes important, as demonstrated in~\cite{ranganath_nayak_contact_1972}. For such a ``heavy'' particle, the oscillation amplitude of the vertical contact displacement $a^*(t^*)$ substantially exceeds the oscillation amplitude of the coordinate $y^*(t^*)$. This means that, because of its greater inertia, the heavy particle tends to deform rather than move. 

Panel (c) in Fig.~\ref{fig:against} shows a fragment of these two curves, highlighting a significant contrast between the maximum and minimum normal loads. The tangential contact displacement $b^*(t^*)$, also shown in the same plot, exhibits advancing and receding stages occurring at low and high compression, respectively. At low compression, the particle fully slides on the substrate with no stick ($s=0$), whereas at high compression slip is only partial ($0<s<c$) or absent ($s=c$). The alternating advancing and receding stages~\textemdash referred here to as \textit{asymmetric sliding}~\textemdash result in a global drift of the particle along the substrate (see the fragment of $x^*(t^*)$ in panel (c) or the entire curve in panel (b)). This phenomenon resembles snake locomotion, as it is based on alternating stick and slip phases occurring at high and low compression, respectively.

After completing a set of numerical experiments, we established two conditions necessary for the existence of drift in the permanent contact case. First, there must be a strong contrast between low and high normal compression (as in the $N^*(t^*)$ or $a^*(t^*)$ curves) that allows full sliding during the advancing stage and suppresses it during the receding stage. Second, the tangential force $T^*$ must maintain a consistent sign during the low compression phase. Under these conditions, total sliding occurs, with its direction determined by the sign of $T^*$.

Indeed, in the full sliding state, the Coulomb friction law reads $T=\sgn(\dot{b})\mu N$. The efficiency of the relative motion is determined by $\dot{b}$. During the low compression phase, $\dot{b}$ should be highly positive or highly negative (see Fig.~\ref{fig:against}(d)); otherwise, drift will not occur. This implies that drift requires a specific phase relationship between forces $N$ and $T$, which can be seen as the system's response to external excitation in terms of wave displacements $u_x(t)$, $u_y(t)$.

\begin{figure*}[htbp]
	\includegraphics[width=\textwidth]{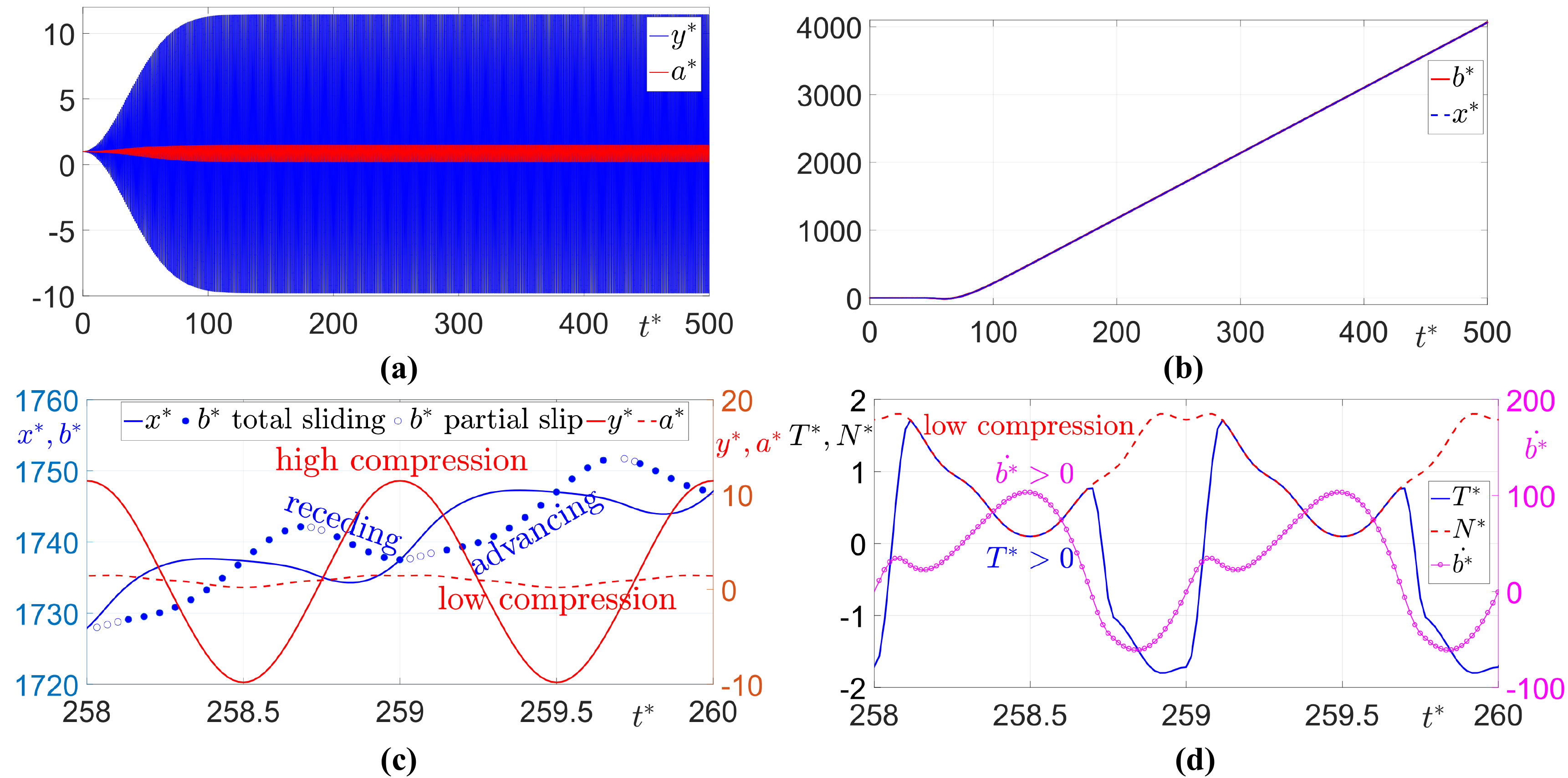}%{Figures/Fig5_with_lgAy-2.7_lgm1.eps}
		\caption{Drift in the wave propagation direction for a deformable particle in permanent contact with a substrate (example for $A_y^*=10^{1}$, $m^*=10^{-2.7}$
        ; point 2 in Fig.~\ref{fig:velocity_MMD})). (a) Normal displacements $a^*(t^*)$ and $y^*(t^*)$. (b) Tangential displacements $b^*(t^*)$ and $x^*(t^*)$. (c) Fragments of the above curves. Advancing in the positive direction occurs during the low compression phase (local minimum of  $a^*(t^*)$) under total sliding. (d) Normal and tangential forces $N^*(t^*)$, $T^*(t^*)$, and horizontal velocity $\dot{b^*}(t^*)$ relative to the substrate. During total sliding, both $T^*$ and $\dot{b^*}$ are positive, resulting in positive drift direction.\label{fig:with}}
\end{figure*}

\paragraph*{\textbf{2) Drift with the wave.}}
In this situation, the particle globally drifts in the direction of the wave propagation via the same mechanism as in the previous case. To observe this regime, parameters $m^*$ and $A_y^*$ must be located in the red zone just below the delimiting line in Fig.~\ref{fig:velocity_MMD}(a) (point 2). The external action ($A_y^*$) must be strong enough to create a pronounced contrast between high and low compression. Here, however, the particle is ``light'' ($m^*<m^*_c$) ), which means it tends to move rather than deform.

This is evident in Fig.~\ref{fig:with}(a), where the oscillation amplitude of $a^*$ is considerably lower than that of $y^*$. Thus, to achieve the required contrast between the loaded and unloaded phases, the wave amplitude $A_y^*$ must be higher than in the previous case (heavy particle).

As previously, horizontal motion relies on the existence of advancing and receding stages (panel~(c)). However, its direction is opposite, as the tangential force $T$ is positive during the low-compression phase (see Fig.~\ref{fig:with}(d)). 

Cases 1 and 2 may be suitable for applications that require fine particle positioning, as the typical horizontal velocity is low, $|v^*|\sim 10^0$.

\begin{figure*}[htbp]
    \includegraphics[width=\textwidth]{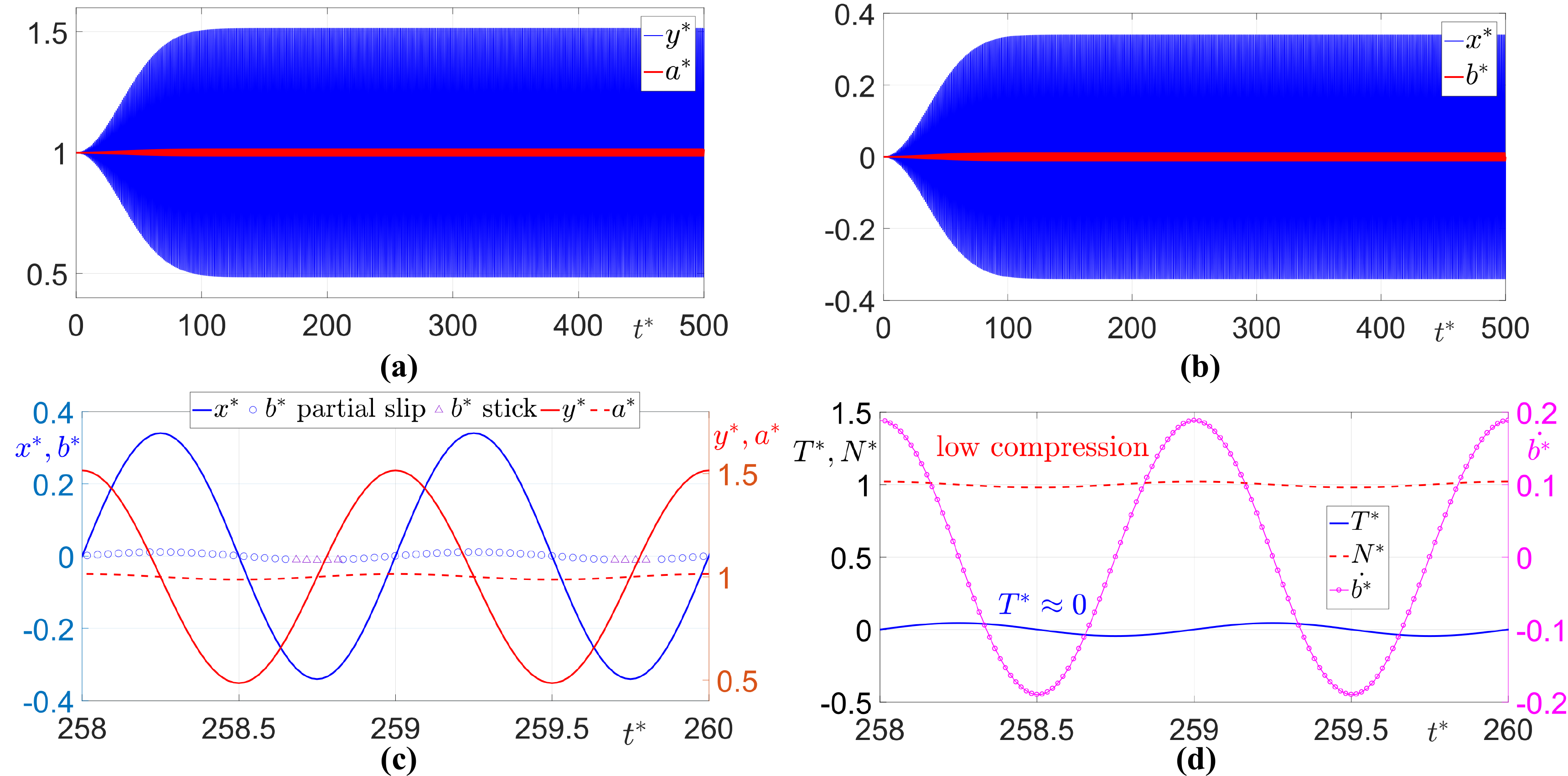}%{Figures/Fig6_weak_lgAy-0.3_lgm-3.eps}
    \caption{No drift for a deformable particle in permanent contact with a substrate ($A_y^*=10^{-0.3}$, $m^*=10^{-3}$; point 3 in Fig.~\ref{fig:velocity_MMD}). (a) Normal displacements $a^*(t^*)$ and $y^*(t^*)$. (b) Tangential displacements $b^*(t^*)$ and $x^*(t^*)$. (c) Fragments of the above curves. (d) Normal and tangential forces $N^*(t^*)$, $T^*(t^*)$, and horizontal velocity $\dot{b^*}(t)$ relative to the substrate. The contrast between high and low load is weak, preventing total sliding. The particle oscillates around its initial position without exhibiting horizontal drift.  \label{fig:weak}}
\end{figure*}

\begin{figure*}[htbp]
		\includegraphics[width=\textwidth]{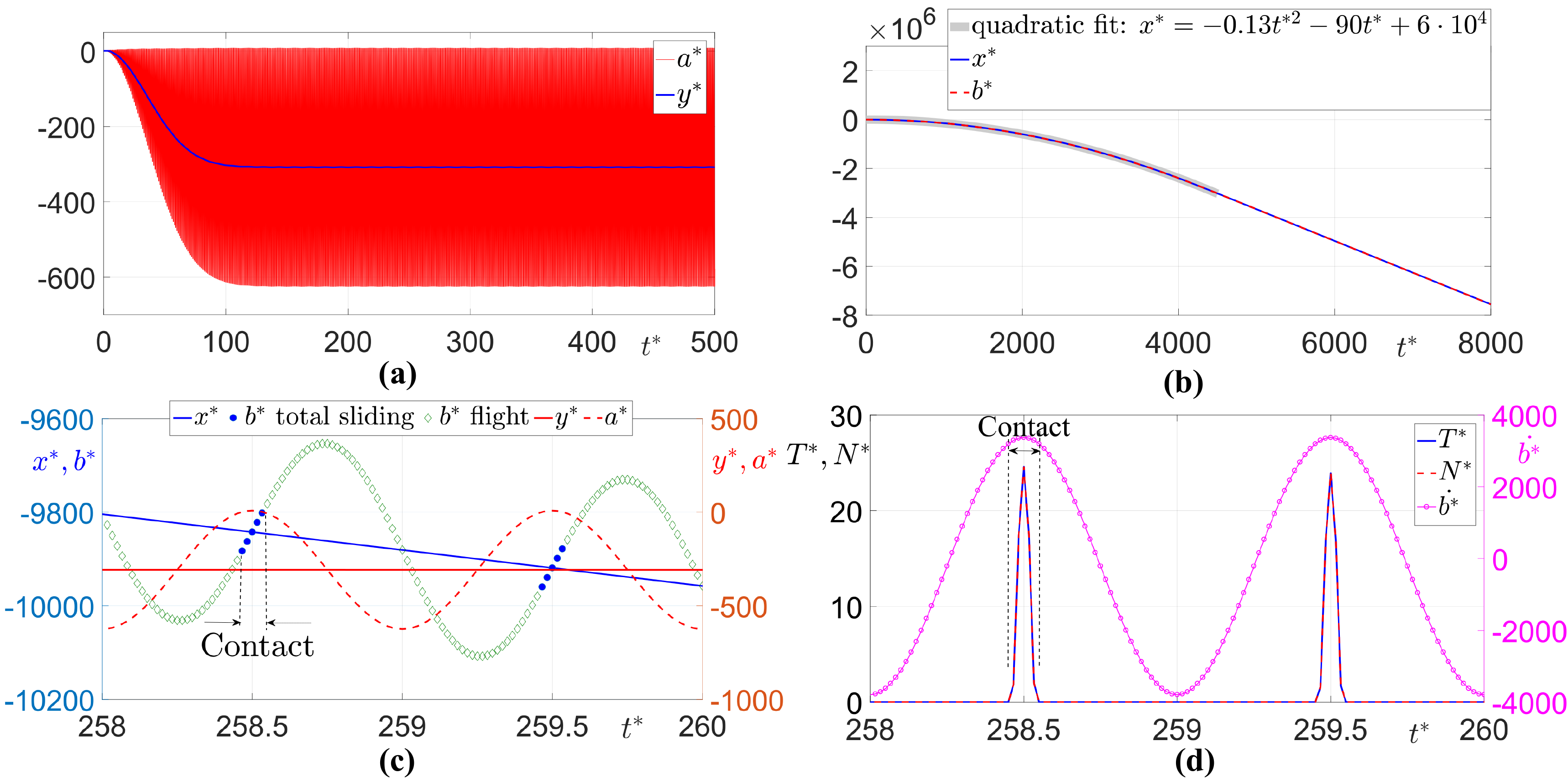}%{Figures/Fig7_hovering_lgAy0_lgm2.5.eps}
		\caption{Synchronous jumping  (example for $A_y^*=10^{2.5}$, $m^*=10^{0}$; point 4 in Fig.~\ref{fig:velocity_MMD}). (a) Vertical displacements in the laboratory frame $y^*(t^*)$ and substrate frame $a^*(t^*)$; note that $y^*(t^*)$ remains nearly constant (hovering effect). (b) Horizontal displacements in the laboratory frame $x^*(t^*)$ and substrate frame $b^*(t^*)$. (c) Fragments of the above curves showing two rebound events (negative $a^*$). (d) Normal and tangential forces $N^*(t^*)$, $T^*(t^*)$ exhibit strong peaks during rebounds, while $\dot{b^*}$ is positive.    \label{fig:hovering}}
\end{figure*}

\paragraph*{\textbf{3) No drift.}}
As shown earlier, certain conditions must be met for drift to occur. For completeness, a no-drift case is presented in Fig.~\ref{fig:weak} (point 3 in Fig.~\ref{fig:velocity_MMD}). In Fig.~\ref{fig:weak}(b), the absence of global drift is evident: both $x^*(t^*)$ and $b^*(t^*)$ oscillate around 0. 

Here, the contrast between highly and weakly loaded loaded stages is insufficient to create the advancing and receding effects. As shown in Fig.~\ref{fig:weak}(c) and (d), both $a^*$ and $N^*$ evolve around their mean value of 1. The total sliding regime necessary for horizontal drift does not appear (see Fig.~\ref{fig:weak}(c)). During the low compression phase, $T^*$ oscillates around 0 with a low amplitude, resulting in no clear drift direction.

To observe drift, the contrast between loading and unloading must be greater than in point 3 of Fig.~\ref{fig:velocity_MMD}. This can be achieved either by increasing the amplitude $A^*_y$ or by increasing the effective mass $m^*$ which effectively means softening the contact system i.e. making it more deformable. In the latter case, the current amplitude ($A_y^*=10^{-0.3}$) may be sufficient to induce drift.

\subsubsection{Deformable particle in the jumping mode\label{sec:jump_def}}
\paragraph*{\textbf{4) Synchronous jumping.}}
In this regime, the particle repeatedly jumps on the substrate with exactly one rebound per wave period, every time occurring in the same phase. In other words, the rebounds are synchronized with the wave (Fig.~\ref{fig:hovering}(c)). At each collision, the upward motion of the substrate imparts vertical momentum to the particle, compensating for the gravitational descent during flight. As a result, there is no net vertical motion; the particle ``hovers'' at a height approximately equal to the vertical wave amplitude (see $|y^*|\approx 300 = A^*_y$ in Fig.\ref{fig:hovering}(a) and (c) in the steady regime).

This vertical regime of motion induces an efficient horizontal drift (see $x^*(t^*)$ in Fig.~\ref{fig:hovering}(b),(c)). The underlying mechanism is related to a velocity mismatch between the particle and the substrate in the horizontal direction during collisions. This mismatch produces a nonzero $\dot{b^*}$ during contact, generating a Coulomb friction force (panel (d)).

Friction, in turn, accelerates the particle horizontally during the total sliding mode (see Fig.~\ref{fig:hovering}(b)). As a result, the particle's horizontal velocity during collisions increases and eventually matches the substrate velocity. This means that finally the acceleration disappears together with $\dot{b}$, and the friction force becomes negligible.

The direction of drift is determined by the sign of the substrate's horizontal velocity during rebound events, which occur when the substrate is near its top position (see $a^*$ in Fig.~\ref{fig:hovering}(c)). For a Rayleigh wave traveling in the positive direction, this velocity is negative, resulting in drift in the opposite (negative) direction.

This motion modality is well suited for applications such as dust cleaning, as the resulting horizontal velocity is high, $v^*\sim 10^3$. However, it requires a relatively large wave amplitude $A_y^*$.

\paragraph*{\textbf{5) Multi-synchronous jumping.}}

\begin{figure*}[htbp]
\includegraphics[width=\textwidth]{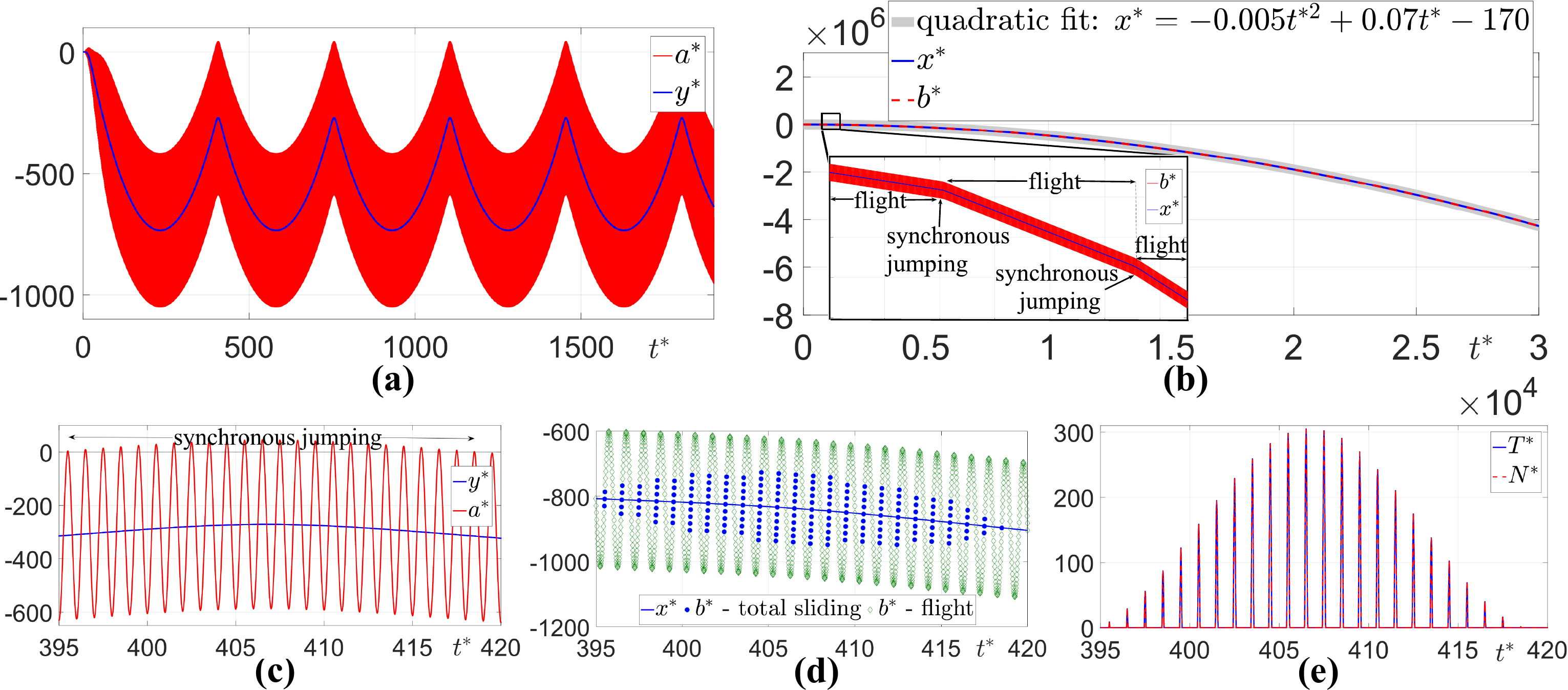}%{Figures/Fig8_multisyn_lgAy2.5_lgm1.5.eps}
\caption{Multi-synchronous jumping (example for $A_y^*=10^{2.5}$, $m^*=10^{1.5}$; point 5 in Fig.~\ref{fig:velocity_MMD}). (a) Vertical displacements in the laboratory frame $y^*(t^*)$ and substrate frame $a^*(t^*)$. (b) Horizontal displacements in the laboratory frame $x^*(t^*)$ and substrate frame $b^*(t^*)$. (c, d) Fragments of normal and tangential solutions, respectively, showing a series of synchronous (i.e., occurring once per period) rebounds. (e) Rebounds correspond to peaks of the contact forces $N^*(t^*)$, $T^*(t^*)$.        \label{fig:multisync}}	
\end{figure*}

\begin{figure*}[htbp]
		\includegraphics[width=\textwidth]{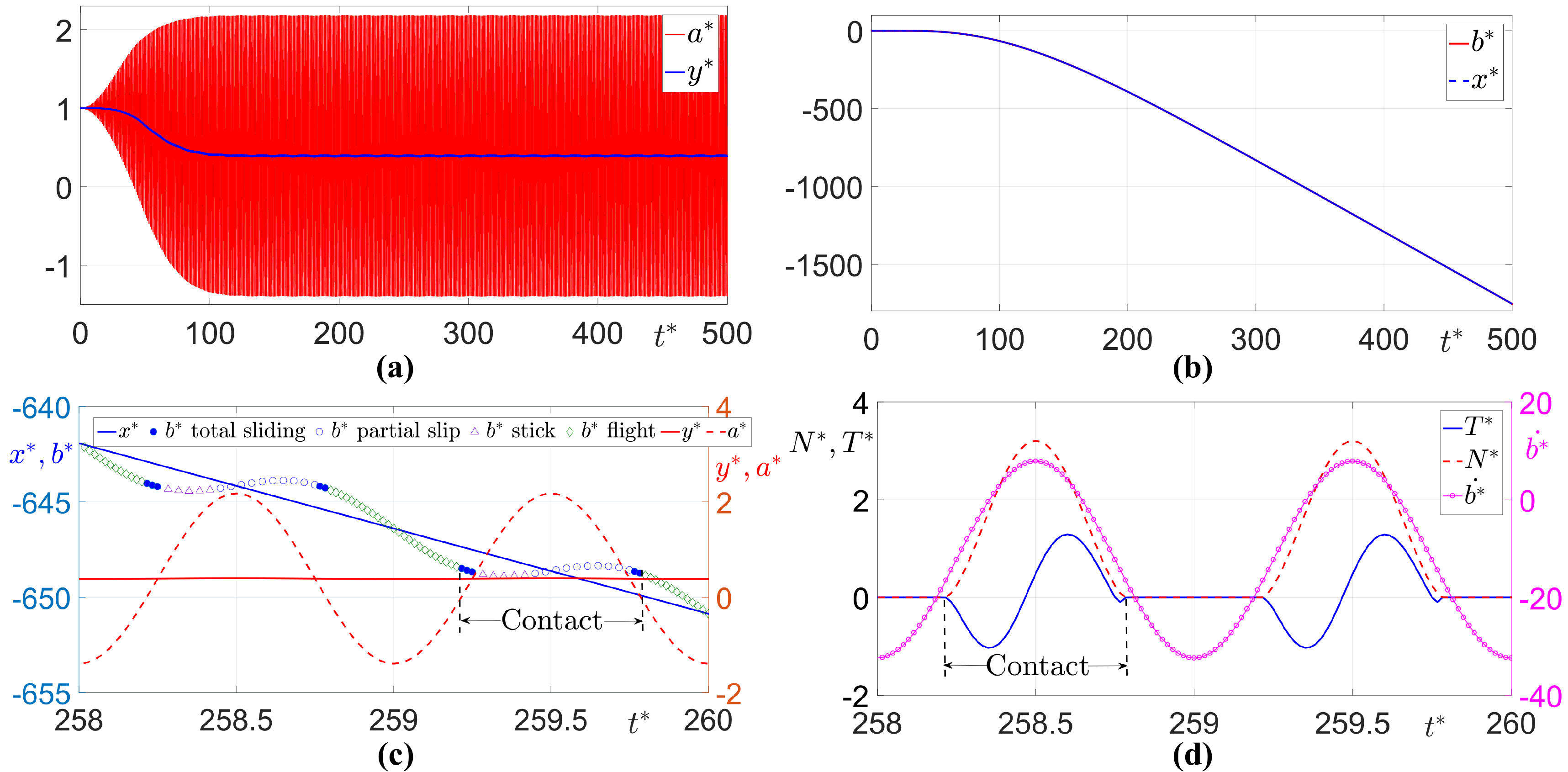}%{Figures/Fig9_mixed_uniform_lgAy0.25_lgm0.8.eps}
		\caption{Combined mechanisms of drift: synchronous jumping and asymmetric sliding (example for $A_y^*=10^{0.25}$, $m^*=10^{0.8}$; point 6 in Fig.~\ref{fig:velocity_MMD}). (a) Vertical displacements in the laboratory frame $y^*(t^*)$ and substrate frame $a^*(t^*)$. (b) Horizontal displacements in the laboratory frame $x^*(t^*)$ and substrate frame $b^*(t^*)$. (c) A two-period segment of $x^*(t^*)$, $b^*(t^*)$, $a^*(t^*)$, $y^*(t^*)$, highlighting the contact phase with stick, partial slip, and total sliding regimes (the latter having negligible effect), followed by the flight phase. (d) Normal and tangential forces $N^*(t^*)$, $T^*(t^*)$, and horizontal velocity $\dot{b^*}(t^*)$ relative to the substrate.\label{fig:sync_jump}}	
\end{figure*}

\begin{figure*}[htbp]
		\includegraphics[width=\textwidth]{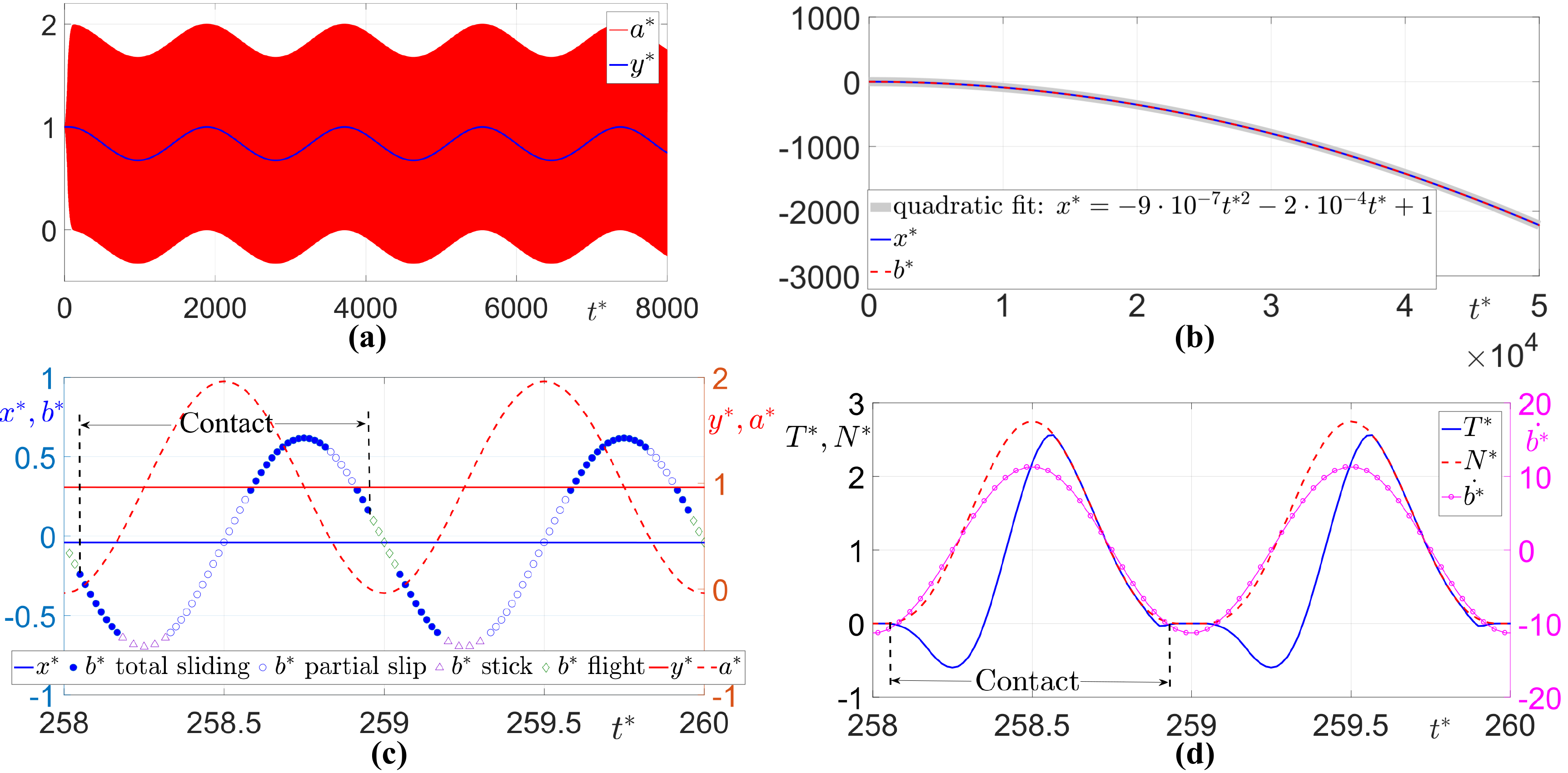}%{Figures/Fig10_mixed_highFreq_lgAy0_lgm5.eps}
		\caption{Combined mechanisms of drift: synchronous jumping and asymmetric sliding (example for $A_y^*=10^{0}$, $m^*=10^{5}$; point 7 in Fig.~\ref{fig:velocity_MMD}). (a) Vertical displacements in the laboratory frame $y^*(t^*)$ and substrate frame $a^*(t^*)$. (b) Horizontal displacements in the laboratory frame $x^*(t^*)$ and substrate frame $b^*(t^*)$. (c) A two-period segment of the above curves, highlighting the contact phase with stick, partial slip, and total sliding regimes, followed by the flight phase. (d) Normal and tangential forces $N^*(t^*)$, $T^*(t^*)$, and horizontal velocity $\dot{b^*}(t^*)$ relative to the substrate.\label{fig:mixed_high_freq}}	
\end{figure*}

\begin{figure*}[htbp]
		\includegraphics[width=\textwidth]{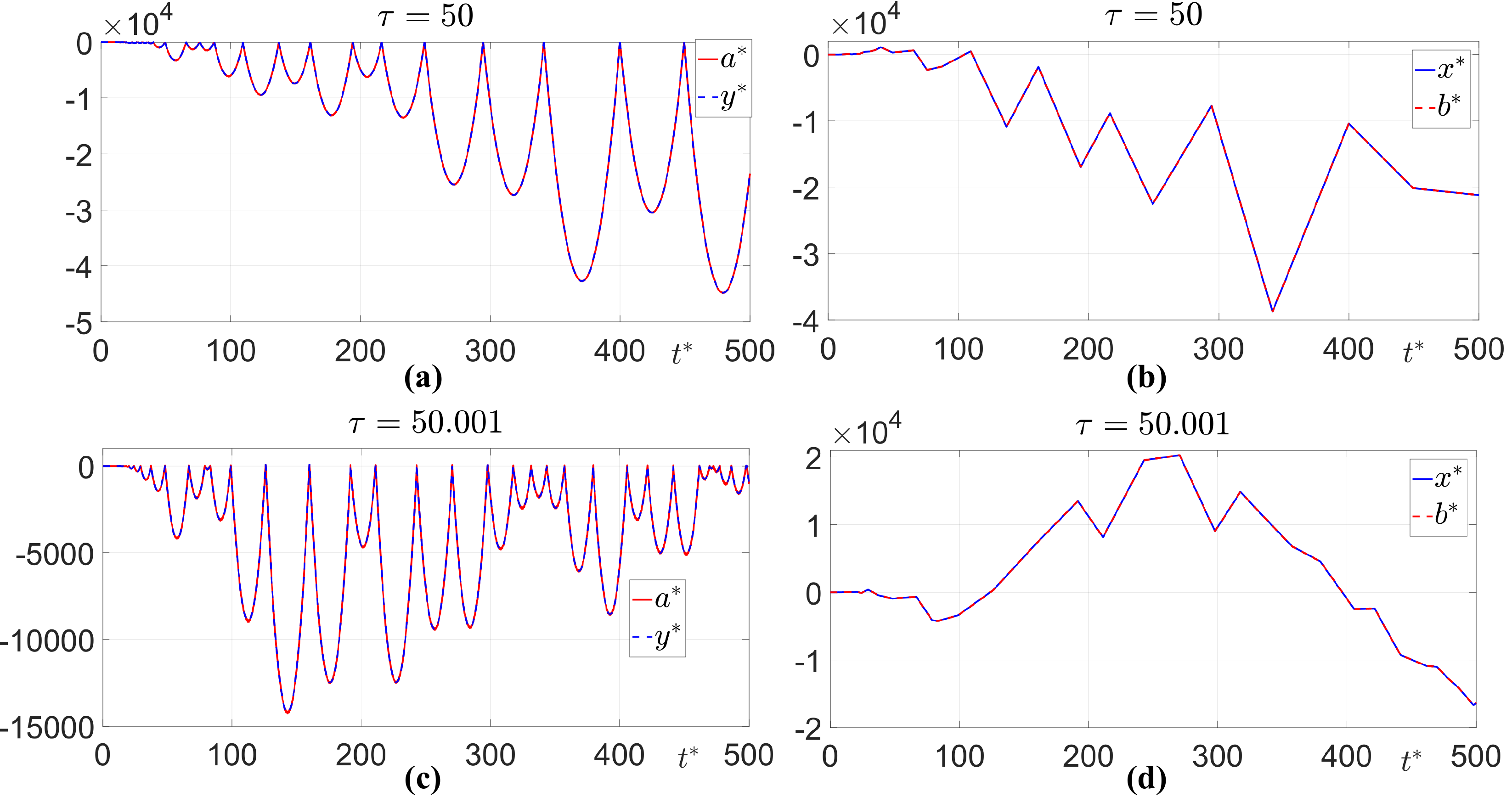}%{Figures/Fig11_chaotic_lgAy1.7_lgm_-2.eps}
		\caption{Chaotic jumping (example for $A_y^*=10^{1.7}$, $m^*=10^{-2}$; point 8 in Fig.~\ref{fig:velocity_MMD}). (a, c) Vertical and (b, d) horizontal displacements under slightly different activation times $\tau$.        \label{fig:chaos}}
\end{figure*}

In the synchronous jumping regime, the momentum gained from the substrate at rebound is compensated by the momentum induced by gravity during flight. This compensation is generally not perfect, except in specially optimized cases where $m^*$ approaches the subharmonic resonance at $m^*=4m_c^*$ dashed line in Fig.~\ref{fig:velocity_MMD}(a)), which separates the chaotic and synchronous zones. For $m^*$ considerably exceeding $4m_c^*$, $y^*$ is not constant on average but oscillates around $A_y^*$ with a low frequency~\textemdash much lower than the excitation frequency $f$.

At higher $m^*$, the difference between the two momenta increases, progressively destroying the hovering effect. A further increase in $m^*$ completely breaks the balance and triggers a curious motion type that can be called multi-synchronous jumping (Fig.~\ref{fig:multisync}). In this regime, the particle bounces several times (23 in the example shown in panels (c) and (d)) once per wave period and then detaches for a longer flight lasting hundreds of periods (panel (a)). Upon landing, the particle again undergoes a series of synchronous jumps once per period and then leaves for another long flight of equal duration, making the process periodic (see the inset in Fig.~\ref{fig:multisync}(b)).

In this regime, as in the case of synchronous jumping, the horizontal motion is initially accelerated when the substrate has a higher horizontal velocity during collisions, then becomes uniform as the velocities equilibrate. However, since the particle spends hundreds of periods in the flight, the contact interaction events are rare. Therefore, the acceleration phase lasts longer, i.e. more time is required to achieve the steady drift (e.g. 2 million wave periods in the example shown in Fig.~\ref{fig:multisync}(b)). Generally, this type of motion is promising for dust cleaning applications and is especially suitable for heavy dust particles.

\paragraph*{\textbf{6-7) Combined mechanisms of drift.}}

So far, we have identified two mechanisms of drift: one related to advancing and receding in contact (asymmetric sliding or snake-like locomotion; see Section~\ref{sec:contact_def} (1), (2)), and another driven by synchronous rebounds that enable horizontal, friction-induced acceleration. In some cases (points 6 and 7 in Fig.~\ref{fig:velocity_MMD}), both mechanisms work together. As previously, jumps occur once per period, allowing consecutive interaction events to produce a coherent net effect. However, these interaction events are more complex and last significantly longer compared to those in hovering or multi-synchronous regimes (approximately 70\%\textendash90\% vs. 6\%\textendash8\% of the period; see Figs.~\ref{fig:sync_jump},~\ref{fig:mixed_high_freq} vs. Fig.~\ref{fig:hovering}). During this extended contact phase, the substrate may move both left and right producing various contact regimes.

In particular, in case 6, during the contact phase (Fig.~\ref{fig:sync_jump}(c)), which lasts about 70\% of the period, the particle undergoes stick, partial slip, and total sliding. However, total sliding is rare, so there is practically no relative motion while in contact. Meanwhile, relatively short jumps (about 30\% of the period) contribute significantly to the particle's drift. As a result, the horizontal motion appears uniform (Fig.~\ref{fig:sync_jump}(b)), since acceleration can only occur during the contact phase.

In case 7, the mechanisms of asymmetric sliding and synchronous jumping are again combined, but their overall effect differs. As in the previous case, the contact phase lasts long (about 90\% of the period), and the particle transitions through all contact regimes. However, unlike before, the influence of total sliding is no longer negligible. It occurs in both directions (see Fig.~\ref{fig:mixed_high_freq}(c)), with motion in the positive direction prevailing. At the same time, a small negative increment in $v_x^*$ develops at the end of each contact phase, and $v_x^*$ remains negative overall (see Fig.~\ref{fig:mixed_high_freq}(b)). As a result, a subsequent jump outweighs the contribution of the contact phase, producing a small net negative displacement (Fig.~\ref{fig:mixed_high_freq}(c)). Moreover, the $v_x^*$-increments generated by asymmetric sliding lead to a globally accelerated motion. This effect did not appear in the previous case due to the negligible contribution of total sliding. 

Both cases 6 and 7 are suitable for practical vibrational transportation. In case 6, a moderate uniform drift is observed, while in case 7, a weak acceleration gradually produces a significant velocity. These drift effects are observed at relatively low excitation amplitudes $A_y^*$ and may be particularly useful in applications where higher amplitudes are not experimentally feasible.

\paragraph*{\textbf{8) Chaotic jumps.}}
A large region in the functional diagram (Fig.~\ref{fig:velocity_MMD}(a), see point 8 as an example) corresponds to a regime of ``chaotic'' jumps, where the solution is unstable despite being governed by the deterministic equations of motion (Eq.~\eqref{eq:motion_norm}). This behavior is illustrated in Fig.~\ref{fig:chaos} for two cases with slightly different ramp function durations, $\tau^*=50$ (a,b) and $\tau^*=50.001$ (c,d), respectively. After a few initial rebounds, the trajectories begin to diverge significantly. Each rebound occurs at a different phase of the Rayleigh wave, therefore all the subsequent increments in the horizontal velocity add incoherently, producing a chaotic effect. Consequently, the specific color of a point in the ``motley'' region of Fig.~\ref{fig:velocity_MMD} is essentially meaningless and may vary depending on the pixel size or time step. As a result, this regime is not suitable for practical vibrational transportation. 

\subsubsection{Dynamics of a material point}\label{sec:mp}

\begin{figure*}[htbp]
\centering
\includegraphics[width=0.9\textwidth]{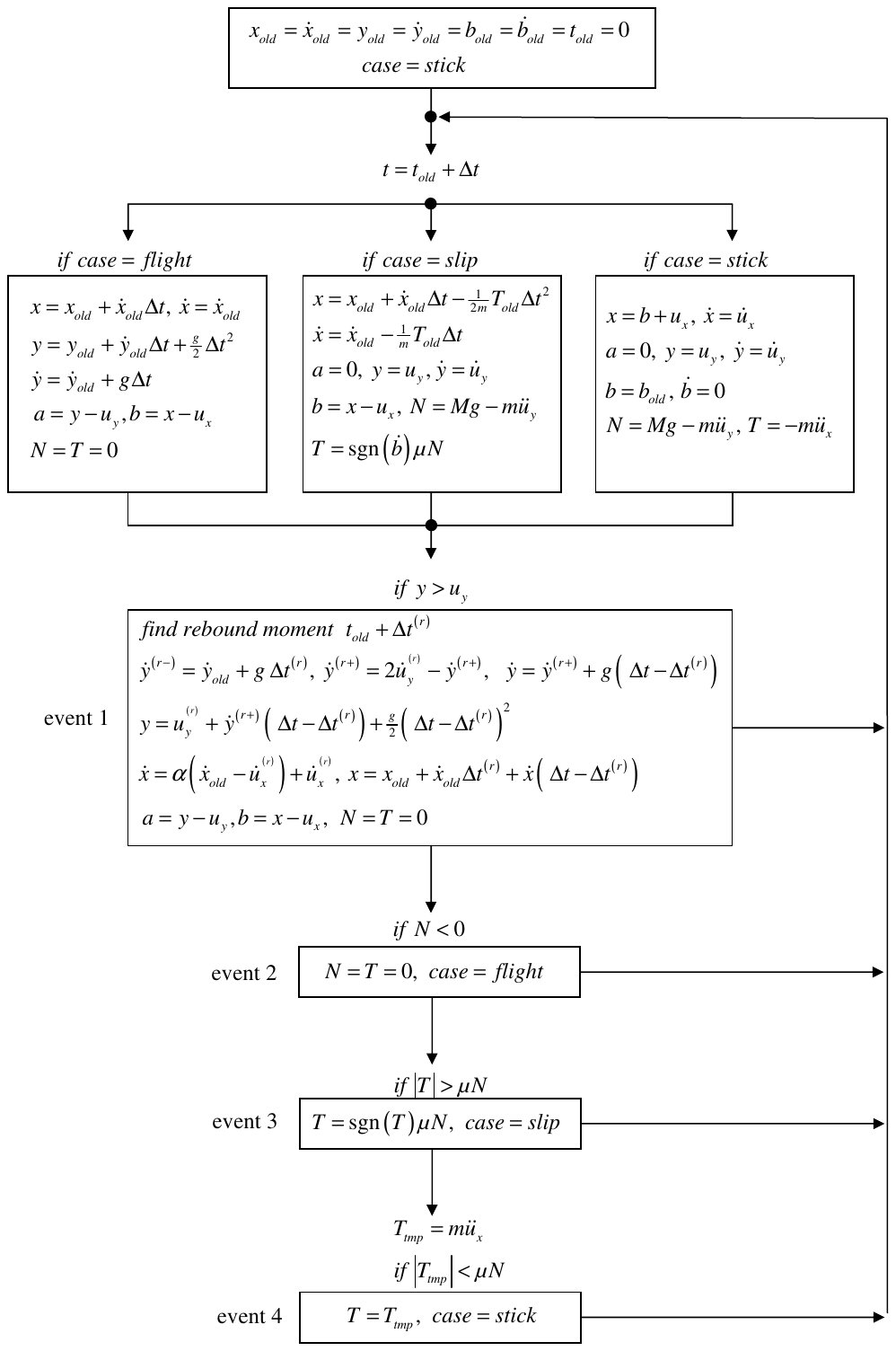}%{Figures/Fig12_mat_point_algo.eps}
	\caption{Material point model: algorithm describing dynamics on a vibrating surface with events 1\textendash4 outlined. The time moment of event 1 is calculated by the bisection method. The time moments of events 2\textendash4 can be refined in the same way.\label{fig:MP_algo}}	
\end{figure*}

\begin{figure*}[htbp]
		\includegraphics[width=\textwidth]
        {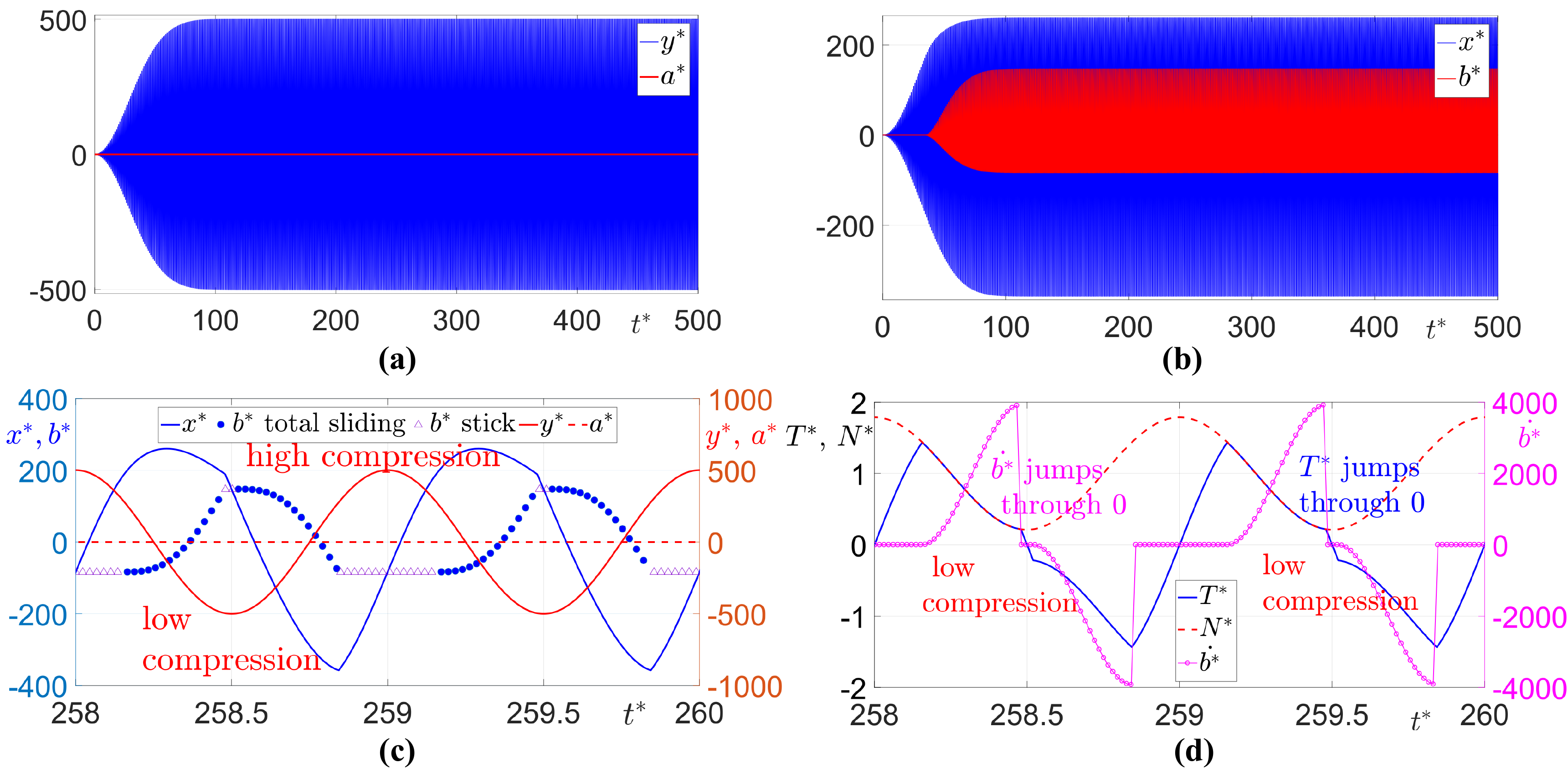}%{Fig13_mp_lgm_-4.4_lgAy_2.7.eps}
        \caption{Symmetric oscillatory sliding of a material point (example for $A_y^*=10^{2.7}$, $m^*=10^{-4.4}$; point 9 in Fig.~\ref{fig:velocity_MMD}). (a) Normal solution. (b) Tangential solution. (c) 2-period fragment of $x^*(t^*)$, $b^*(t^*)$, $a^*(t^*)$, $y^*(t^*)$. Advancing in the positive and negative directions fully compensate each other. (d) Normal and tangential forces $N^*(t^*)$, $T^*(t^*)$, and horizontal velocity $\dot{b^*}(t^*)$ relative to the substrate. The latter experiences sudden drops occurring when the transition to stick (event 4 in Fig.~\ref{fig:MP_algo}) takes place.\label{fig:with_MP}}	
\end{figure*}

\begin{figure*}[htbp]
	\includegraphics[width=\textwidth]{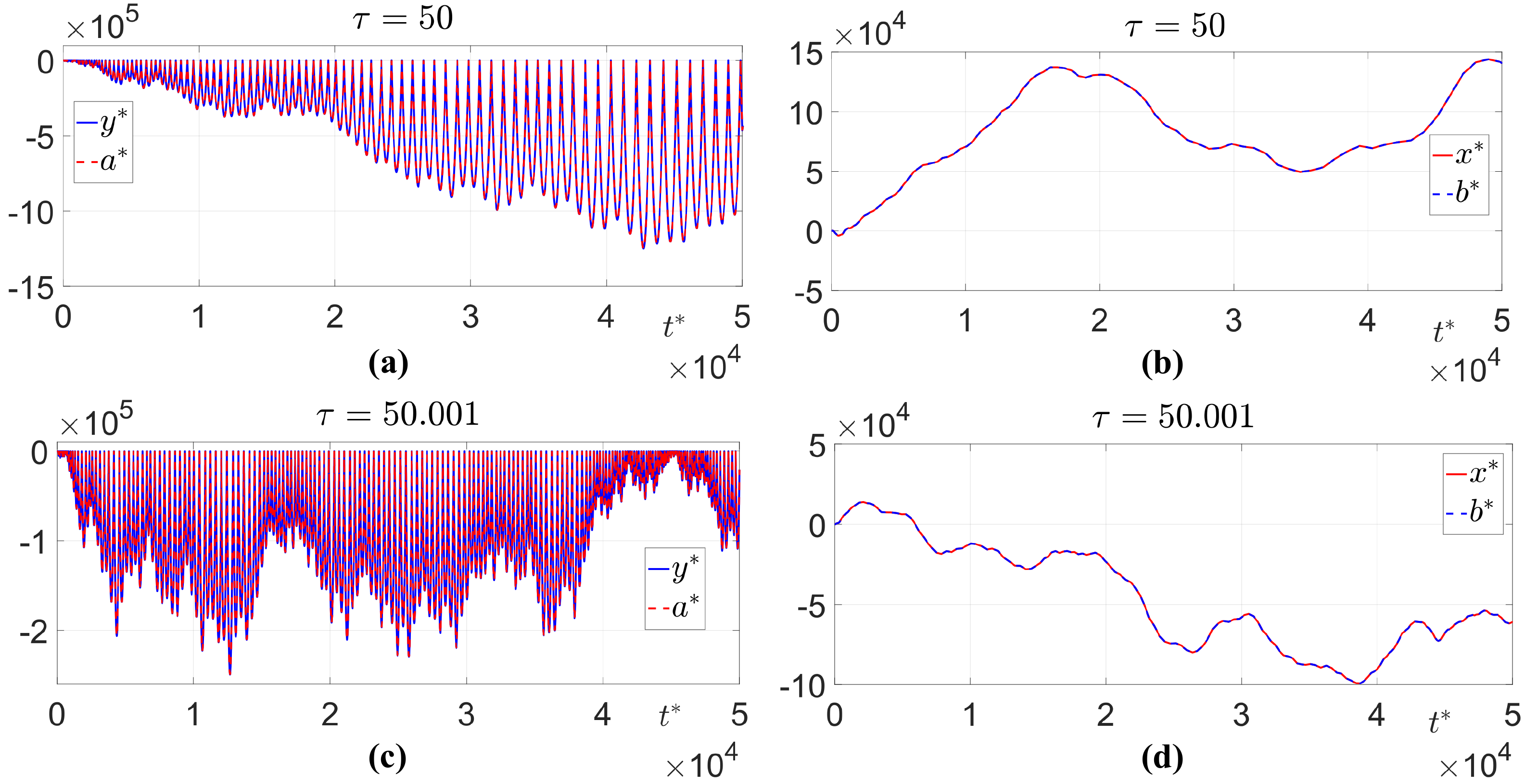}%{Figures/Fig14_mp_chaotic_lgAy1.2_lgm_-1.eps}
		\caption{Unstable jumping of a material point (example for $A_y^*=10^{1.2}$, $m^*=10^{-1}$, point 10 in Fig.~\ref{fig:velocity_MMD}). (a, c)  Vertical and (b, d) horizontal displacements with slightly different activation time $\tau$.\label{fig:chaos_MP}}
\end{figure*}

In contrast to the above results, this section is based on a more traditional point-mass approach and does not strive for originality. Its objective is to demonstrate the difference between the dynamics of deformable and rigid particles, highlighting the importance of accounting for deformability. Fig.~\ref{fig:MP_algo} illustrates the implementation details of the material point model. 

For a point mass, partial slip regime does not make sense; only sticking, sliding, and flight may occur. At each time step, the dynamical variables are updated using known solutions for these regimes. However, doing so, one can face an anomalous event that produces a correction to the solution (see events 1\textendash4 in Fig.~\ref{fig:MP_algo}). Firstly, if during a flight the particle ``penetrates'' under the substrate surface ($y>u_y$, event 1), the solution should be corrected by taking into account a rebound event. Secondly, $N<0$ during the contact phase actually means particle's detachment (event 2). Thirdly, during the stick regime the tangential force calculated using the appropriate solution may exceed the Coulomb threshold, thus producing the transition to the sliding regime (event 3). Finally, if during the slip regime the value of $|m\ddot{u}_x|$ becomes lower than $\mu N$, then the external force in the substrate reference frame is not sufficient to maintain sliding according to the Coulomb's friction law, and the particle sticks (event 4). For better precision, the moments of the anomalous events of type 1\textendash4 are numerically calculated with the bisection method.

Applying the algorithm in Fig.~\ref{fig:MP_algo} to calculate the point mass' coordinates, we found three distinct regimes illustrated in Fig.~\ref{fig:velocity_MMD}(b): no motion, sliding in both directions with no drift (point 9), and chaotic jumping (point 10). In diagram Fig.~\ref{fig:velocity_MMD}(b), the three corresponding areas are separated by two curves (straight lines in the double logarithmic space). The numerical results for these curves are in accordance with their analytical definitions obtained in~\cite{verma_particle_2013}:

\begin{equation}
m^*=\sqrt{\frac{\mu^2}{\mu^2+r^2(\nu)}}\frac{1}{4\pi^2A_y^*}\textrm{,}
\label{eq:sticking_sliding_condition}
\end{equation}
and
\begin{equation}
m^*=\frac{1}{4\pi^2A_y^*}\textrm{.}
\label{eq:sliding_jumping_condition}
\end{equation}

The authors of~\cite{verma_particle_2013} also provide solutions in the case of continuous sliding in the band
between the lines Eq.~\eqref{eq:sticking_sliding_condition} and Eq.~\eqref{eq:sliding_jumping_condition}. In other words, they neglect possible transitions to stick that may occur if the external force in the Coulomb friction law becomes lower than the threshold (event 4 in Fig.~\ref{fig:MP_algo}). In that situation, drift with the wave and against it is predicted. However, an accurate account for this possibility considerably changes the particle's dynamics. 

\paragraph*{\textbf{9) Symmetric oscillatory sliding.}}
Our numerical modeling results indicate that for $m^*$ in between the lines Eq.~\eqref{eq:sticking_sliding_condition} and Eq.~\eqref{eq:sliding_jumping_condition} the particle is sliding continuously. Unlike in~\cite{verma_particle_2013}, positive and negative displacements of a particle fully compensate for each other during each wave period (Fig.~\ref{fig:with_MP}(c)). Thus, asymmetric sliding occurring for a deformable particle becomes symmetric and no drift appears.

Transitions to stick represented as sudden drops of $\dot{b}$ in Fig.~\ref{fig:with_MP}(d) deserve an additional comment. In order to instantly stop a particle of mass $m$ moving with velocity $v_x$ at a moment $t_0$, an infinite force $mv_x\delta(t-t_0)$ should be applied (here $\delta$ is the Dirac delta function). In the absence of other interactions, it can only happen due to friction. However, the Coulomb friction law does not provide such an opportunity because of its quasi-static nature. This means that the Coulomb friction model applied to a material point dynamics should be used with caution.

\paragraph*{\textbf{10) Chaotic jumping.}}
First of all, it should be noted that dynamics of a point mass in the jumping regime is not determined by the Coulomb friction law all the time, but only before the first detachment. To describe subsequent rebounds, one has to know a link between particle velocities before and after collisions. Here we consider fully elastic vertical collisions, and therefore the link for the normal velocities is known. However, for the tangential velocity components it is not. This link would exist, for instance, if during collisions full sliding was guaranteed. Indeed, in that case one would be able to calculate the tangential momentum change through the normal one by using the relation $|T|=\mu N$. But, in a general case, no physical link between contact forces exists during instantaneous rebounds of a material point. Alternatively, tangential interactions can also be considered conservative, but this means that friction is absent. In this case, the horizontal particle velocity is not affected by the Rayleigh wave and remains constant.

A usual way to overcome this difficulty is to introduce a horizontal restitution coefficient $\alpha_t$ defined as the ratio between the tangential velocities after and before the rebound for a stationary substrate. Note that for a deformable particle described by an MMD-type model (see Section~\ref{sec:MMD}), such an artificial assumption is not required; the system is entirely governed by the Coulomb friction law. Here we suppose that $\alpha_t<1$ while the respective normal restitution coefficient $\alpha_n=1$ for fully elastic vertical collisions. 

An important feature of a point mass in the jumping regime is that the tangential motion is unstable (see Fig.~\ref{fig:velocity_MMD}(b)) for any $\alpha_t<1$. This occurs since the vertical motion is also unstable, at least for the conservative normal interaction. The chaotic character of a vertical conservative motion of a material point on a vibrating substrate has been analytically demonstrated by the authors of~\cite{luck_bouncing_1993}. Indeed, it was shown that at $\alpha_n=1$ a stable motion is possible only for

\begin{equation}
m^*<\frac{1}{4\pi^2A_y^*}\textrm{,}
\label{eq:stability}
\end{equation}

(see Eq.~2.14 with $\alpha=1$ in there), for any initial conditions. However, the analytical results~\cite{verma_particle_2013} obtained in the stationary excitation case similar to the one considered here, indicate that in the situation Eq.~\eqref{eq:stability} the particle does not jump at all. This means that, in our case, stable jumping solutions for a material point do not exist.

Our numerical computations confirm the above statement (see Fig.~\ref{fig:chaos_MP}). Indeed, a slight change in the ramp function activation time produces a completely different solution, similarly to the case of chaotic jumping of a deformable particle (see case 8 in Section~\ref{sec:jump_def}).

\section{Discussion\label{sec:conclusions}}
In this paper, we discuss possibilities of tangential drift of a deformable particle posed on a substrate in which a traveling Rayleigh wave is excited. Depending on values of two governing parameters $A_y^*$ and $m^*$, two situations can be observed. The particle can stay in permanent contact with the substrate or experience multiple jumps. It is demonstrated that mechanisms of directed motion are different in these cases.

In permanent contact, the drift mechanism resembles snake locomotion and is called here asymmetric sliding. This means that sliding can occur in both directions during one period, but one of the two sliding paths is longer. An obvious condition for the asymmetry is a high contrast between high and low vertical load; at weaker loads sliding is more efficient. In addition, some phase relationship between the normal and tangential forces is required. If the tangential force changes sign at the lowest compression instead of being directed in one of the two directions, the net horizontal motion is absent. Efficient drift occurs if the difference in the advancing and receding sliding paths is significant. Note that sliding may not occur during the receding stage, whereas it plays a crucial role in generating drift during the advancing stage.

In the jumping mode, the existence of horizontal drift is fully determined by the character of normal motion, which can be synchronized with the wave or chaotic. In the former case, the consecutive particle rebounds occur in the same wave phase once per period, and particle's horizontal momentum increments produced by the friction force are summed up in a coherent way, thus resulting in a drift. In contrast, in the latter situation, rebounds happen in different wave phases, and therefore the friction-induced momentum increments are summed up incoherently.  Hence, the tangential particle motion is chaotic as well, and no drift is observed. In addition, the particle trajectories are poorly determined by initial conditions.

The original deformable particle model is implemented here with the use of the Method of Memory Diagrams which originates from the Hertz-Mindlin solution. We also implemented a traditional material point model moving on a rigid substrate. The results are essentially different; the material point in the jumping mode always moves chaotically and never synchronizes with the wave, at least for fully conservative vertical redounds. Besides, we have found that in the state of continuous contact, during each wave period, the material point covers the same path in the direction of the wave and against it. Generally, the material point model should be used with caution when applied to dynamic problems with Coulomb friction. Indeed, Coulomb friction produces instantaneous transitions to stick when the external force falls below the threshold $\mu N$ but does not provide a proper force that can be responsible for the instantaneous stops. In the jumping regime, the Coulomb friction law alone can not explain tangential motion; other models such as inelastic tangential restitution coefficient should be used. It is worth noting that the deformable contact model based on the MMD or other solutions of this class is free from these disadvantages.

Generally, the deformable particle exhibits a wider variety of directional motion regimes. In addition to asymmetric sliding and synchronous jumping against the wave, combined motion types can appear. The contact phase can last for a considerable part of the period (e.g. 30\%\textendash90\%) and can comprise all contact regimes (stick, partial slip, total sliding), which is not possible for a material point always capable of instantaneous rebound only.

In many cases, an accelerated hopping motion against the wave is observed; the acceleration disappears when the drift velocity reaches the horizontal substrate velocity at take-off moments. Accelerated motion has been experimentally observed in~\cite{kurosawa_friction_1996,morita_simulation_1999,bao_transport_2009,dunst_vibration-assisted_2018}. Our first experimental results (paper in preparation) also indicate initially accelerated drift in the direction against the wave, followed by a uniform motion.

These experimental observations are not reproduced by the point-mass model, at least in the case of conservative vertical collisions. The rigid character of the particle-substrate interaction produces other peculiar effects. For instance, Fig.~\ref{fig:vertical}(b) shows that a point mass jumps on a substrate even for an infinitely small wave amplitude, while a deformable particle does not show this feature (jumping zone is bounded at small amplitudes in Fig.~\ref{fig:vertical}(a)). 

However, the use of the deformable particle model is associated with difficulties that do not arise for the material point. Deformability means a finite size of the particle, which in turn activates rolling, tilting, and torsion. Such degrees of freedom do not exist for a point. This means that, in order to perform a detailed theory and experiment comparison, these motion types should be either disabled experimentally, or included into the model. 

Nevertheless, the results of the deformable particle model (functional diagram Fig.~\ref{fig:velocity_MMD}(a)) represent a useful guideline for experiments on particle manipulation or vibrational transportation. A point with coordinates ($A_y^*$, $m^*$) corresponds to a set of wave and particle parameters for which drift with a certain velocity is expected. Velocity in Fig.~\ref{fig:velocity_MMD}(a) varies in a broad range between $10^0-10^1$ for the asymmetric sliding regime and is of order of $10^3$ for synchronous jumping. In practice, the high-speed synchronous jumping observable in a wide range of system parameters (blue area in Fig.~\ref{fig:velocity_MMD}(a)) can be used for self-cleaning of vibrating plate-like structures. Asymmetric sliding is considerably slower but can occur in both directions. This can be of use for precise particle manipulation. Note, however, that asymmetric sliding happens in the narrow band of control parameters ($A_y^*$, $m^*$) making it challenging to observe experimentally. 

A value of the friction coefficient has moderate impact on horizontal motion. We have observed that lower friction coefficient broadens the band of asymmetric sliding in Fig.~\ref{fig:velocity_MMD}(a) and generally increases its speed. At the same time, in the case of synchronous jumping, lower $\mu$ extends the accelerated motion duration in Fig.~\ref{fig:hovering}(b), since the driving friction force becomes lower in this situation. Conversely, at higher values of $\mu$, the final constant velocity is achieved faster. Note that the value of this constant velocity is not affected by $\mu$ and is approximately equal to the horizontal substrate velocity during collision events (more precisely, to the take-off velocity). 

There is at least one important question that remains unanswered in the framework of this paper. Our numerical results indicate zones of synchronous and chaotic jumping, but do not explain their configurations. For instance, in Fig.~\ref{fig:vertical}(a), these zones are delimited by a value of $m^* \approx 4m^*_c$ that corresponds to the subharmonic Hertzian resonance. To explain this fact, a stability analysis similar to~\cite{luck_bouncing_1993} can be of use. However, to do that, one needs, for instance, a link between vertical landing and take-off velocities and the collision time. For a point mass, this problem does not appear, since the collisions are instantaneous, but here some information coming from the specific interaction law must be present. An analytical progress related to a jumping Hertzian particle description is limited. As for the approximate methods for investigation of nonlinear systems such as the harmonic balance~\cite{nayfeh_response_1983} or the multiple scale method~\cite{hess_normal_1991}, they require a small parameter which is not obvious to define far from the fundamental Hertz resonance. Another prospective method of vibrational dynamics analysis at frequencies higher than the system resonance is the method of direct separation of motions~\cite{blekhman_effects_2016} that does not require a small parameter.

\begin{acknowledgments}
M.~Terzi would like to thank \textit{Institut d’Acoustique - Graduate School} (IA-GS) of Le Mans University for funding her postdoctoral research. J.~Ghesquière acknowledges \textit{Hauts-de-France} region, \textit{Fonds Européen de Développement Régional} (FEDER) and the University of Lille for funding his PhD studies.    
\end{acknowledgments}

\bibliographystyle{apsrev4-2.bst}
\bibliography{VibTransport_PhysRevE}

@inbook{blekhman_basic_1994,
	address = {Singapore},
	chapter = {9},
	booktitle = {Vibrational mechanics: Nonlinear Dynamic Effects, General Approach, Applications},
	publisher = {World Scientific},
	author = {Blekhman, I.I.},
	year = {2000},
	pages = {233--261},
    doi = {10.1142/4116},
}

@article{kurosawa_friction_1996,
	title = {Friction drive surface acoustic wave motor},
	volume = {34},
	copyright = {https://www.elsevier.com/tdm/userlicense/1.0/},
	issn = {0041624X},
	url = {https://linkinghub.elsevier.com/retrieve/pii/0041624X96817811},
	doi = {10.1016/0041-624X(96)81781-1},
	number = {2-5},
	urldate = {2025-02-06},
	journal = {Ultrasonics},
	author = {Kurosawa, Minoru and Takahashi, Masakazu and Higuchi, Toshiro},
	month = jun,
	year = {1996},
	pages = {243--246},
}

@article{halev_bouncing_2018,
	title = {Bouncing ball on a vibrating periodic surface},
	volume = {28},
	issn = {1054-1500, 1089-7682},
	url = {https://pubs.aip.org/cha/article/28/9/096103/650371/Bouncing-ball-on-a-vibrating-periodic-surface},
	doi = {10.1063/1.5023397},
	number = {9},
	urldate = {2025-02-06},
	journal = {Chaos: An Interdisciplinary Journal of Nonlinear Science},
	author = {Halev, Avishai and Harris, Daniel M.},
	month = sep,
	year = {2018},
	pages = {096103},
}

@article{luck_bouncing_1993,
	title = {Bouncing ball with a finite restitution: {Chattering}, locking, and chaos},
	volume = {48},
	copyright = {http://link.aps.org/licenses/aps-default-license},
	issn = {1063-651X, 1095-3787},
	shorttitle = {Bouncing ball with a finite restitution},
	url = {https://link.aps.org/doi/10.1103/PhysRevE.48.3988},
	doi = {10.1103/PhysRevE.48.3988},
	number = {5},
	urldate = {2025-02-06},
	journal = {Physical Review E},
	author = {Luck, J. M. and Mehta, Anita},
	month = nov,
	year = {1993},
	pages = {3988--3997},
}

@article{verma_particle_2013,
	title = {Particle current on flexible surfaces excited by harmonic waves},
	volume = {88},
	copyright = {http://link.aps.org/licenses/aps-default-license},
	issn = {1539-3755, 1550-2376},
	url = {https://link.aps.org/doi/10.1103/PhysRevE.88.052915},
	doi = {10.1103/PhysRevE.88.052915},
	number = {5},
	urldate = {2025-02-06},
	journal = {Physical Review E},
	author = {Verma, Neeta and DasGupta, Anirvan},
	month = nov,
	year = {2013},
	pages = {052915},
}

@article{ranganath_nayak_contact_1972,
	title = {Contact vibrations},
	volume = {22},
	copyright = {https://www.elsevier.com/tdm/userlicense/1.0/},
	issn = {0022460X},
	url = {https://linkinghub.elsevier.com/retrieve/pii/0022460X7290168X},
	doi = {10.1016/0022-460X(72)90168-X},
	number = {3},
	urldate = {2025-02-06},
	journal = {Journal of Sound and Vibration},
	author = {Ranganath Nayak, P.},
	month = jun,
	year = {1972},
	pages = {297--322},
}

@article{perret-liaudet_resonance_1998,
	title = {Résonance surharmonique d'ordre deux dans un contact sphère-plan (Second-order superharmonic resonance in a sphere-plane contact)},
	volume = {326},
	copyright = {https://www.elsevier.com/tdm/userlicense/1.0/},
	issn = {12874620},
	url = {https://linkinghub.elsevier.com/retrieve/pii/S1251806999800283},
	doi = {10.1016/S1251-8069(99)80028-3},
	number = {12},
	urldate = {2025-02-06},
	journal = {Comptes Rendus de l'Académie des Sciences - Series IIB - Mechanics-Physics-Astronomy},
	author = {Perret-Liaudet, Joël},
	month = jan,
	year = {1998},
	pages = {787--792},
    note = {(in French)},
}

@article{luo_dynamics_1996,
	title = {The dynamics of a bouncing ball with a sinusoidally vibrating table revisited},
	volume = {10},
	copyright = {http://www.springer.com/tdm},
	issn = {0924-090X, 1573-269X},
	url = {http://link.springer.com/10.1007/BF00114795},
	doi = {10.1007/BF00114795},
	number = {1},
	urldate = {2025-02-06},
	journal = {Nonlinear Dynamics},
	author = {Luo, Albert C. J. and Han, Ray P. S.},
	month = may,
	year = {1996},
	pages = {1--18},
}

@article{nayfeh_response_1983,
	title = {The response of single degree of freedom systems with quadratic and cubic non-linearities to a subharmonic excitation},
	volume = {89},
	copyright = {https://www.elsevier.com/tdm/userlicense/1.0/},
	issn = {0022460X},
	url = {https://linkinghub.elsevier.com/retrieve/pii/0022460X83903474},
	doi = {10.1016/0022-460X(83)90347-4},
	number = {4},
	urldate = {2025-02-06},
	journal = {Journal of Sound and Vibration},
	author = {Nayfeh, A.H.},
	month = aug,
	year = {1983},
	pages = {457--470},
}

@article{hess_normal_1991,
	title = {Normal {Vibrations} and {Friction} {Under} {Harmonic} {Loads}: {Part} {I}—{Hertzian} {Contacts}},
	volume = {113},
	issn = {0742-4787, 1528-8897},
	shorttitle = {Normal {Vibrations} and {Friction} {Under} {Harmonic} {Loads}},
	url = {https://asmedigitalcollection.asme.org/tribology/article/113/1/80/437681/Normal-Vibrations-and-Friction-Under-Harmonic},
	doi = {10.1115/1.2920607},
	number = {1},
	urldate = {2025-02-06},
	journal = {Journal of Tribology},
	author = {Hess, D. P. and Soom, A.},
	month = jan,
	year = {1991},
	pages = {80--86},
}

@book{blekhman_vibrational_1964,
	address = {Moscow},
	title = {Vibratsionnoe peremeshchenie (Vibrational Transportation)},
	publisher = {Nauka Press},
	author = {Blekhman, I. I. and Dzhanelydze, G. Yu.},
	year = {1964},
    note = {(in Russian)},
}

@inproceedings{bao_transport_2009,
	address = {San Diego, California, USA},
	title = {Transport powder and liquid samples by surface acoustic waves},
	volume = {7291},
	url = {http://proceedings.spiedigitallibrary.org/proceeding.aspx?doi=10.1117/12.815387},
	doi = {10.1117/12.815387},
	urldate = {2025-02-06},
	booktitle = {Nanosensors, {Biosensors}, and {Info}-{Tech} {Sensors} and {Systems}},
	author = {Bao, Xiaoqi and Bar-Cohen, Yoseph and Sherrit, Stewart and Badescu, Mircea and Louyeh, Sahar},
	editor = {Varadan, Vijay K.},
	month = mar,
	year = {2009},
	pages = {162--168},
}

@article{abd-elhady_new_2024,
	title = {A new cleaning method for solar panels inspired from the natural vibrations of tree branches and leaves},
	volume = {14},
	issn = {2045-2322},
	url = {https://www.nature.com/articles/s41598-024-68215-y},
	doi = {10.1038/s41598-024-68215-y},
	number = {1},
	urldate = {2025-02-06},
	journal = {Scientific Reports},
	author = {Abd-Elhady, M. S. and Rana, Adil and Elsebaaie, M. A. and Kandil, H. A.},
	month = aug,
	year = {2024},
	pages = {18138},
}

@article{dunst_vibration-assisted_2018,
	title = {Vibration-{Assisted} {Handling} of {Dry} {Fine} {Powders}},
	volume = {7},
	copyright = {https://creativecommons.org/licenses/by/4.0/},
	issn = {2076-0825},
	url = {https://www.mdpi.com/2076-0825/7/2/18},
	doi = {10.3390/act7020018},
	number = {2},
	urldate = {2025-02-06},
	journal = {Actuators},
	author = {Dunst, Paul and Bornmann, Peter and Hemsel, Tobias and Sextro, Walter},
	month = apr,
	year = {2018},
	pages = {18},
}

@article{perret-liaudet_response_2006,
	title = {Response of an impacting {Hertzian} contact to an order-2 subharmonic excitation : {Theory} and experiments},
	volume = {296},
	copyright = {https://www.elsevier.com/tdm/userlicense/1.0/},
	issn = {0022460X},
	shorttitle = {Response of an impacting {Hertzian} contact to an order-2 subharmonic excitation},
	url = {https://linkinghub.elsevier.com/retrieve/pii/S0022460X06002288},
	doi = {10.1016/j.jsv.2006.03.004},
	number = {1-2},
	urldate = {2025-02-06},
	journal = {Journal of Sound and Vibration},
	author = {Perret-Liaudet, J. and Rigaud, E.},
	month = sep,
	year = {2006},
	pages = {319--333},
}

@book{galin_contact_1961,
	address = {North Carolina State College},
	title = {Contact {Problems} in the {Theory} of {Elasticity}.},
	publisher = {Raleigh},
	author = {Galin, L.A.},
	year = {1961},
}

@article{brunet_droplet_2010,
	title = {Droplet displacements and oscillations induced by ultrasonic surface acoustic waves: {A} quantitative study},
	volume = {81},
	copyright = {http://link.aps.org/licenses/aps-default-license},
	issn = {1539-3755, 1550-2376},
	shorttitle = {Droplet displacements and oscillations induced by ultrasonic surface acoustic waves},
	url = {https://link.aps.org/doi/10.1103/PhysRevE.81.036315},
	doi = {10.1103/PhysRevE.81.036315},
	number = {3},
	urldate = {2025-02-06},
	journal = {Physical Review E},
	author = {Brunet, P. and Baudoin, M. and Matar, O. Bou and Zoueshtiagh, F.},
	month = mar,
	year = {2010},
	pages = {036315},
}

@article{erdesz_experimental_1988,
	title = {Experimental study on the vibrational transport of bulk solids},
	volume = {55},
	copyright = {https://www.elsevier.com/tdm/userlicense/1.0/},
	issn = {00325910},
	url = {https://linkinghub.elsevier.com/retrieve/pii/0032591088800913},
	doi = {10.1016/0032-5910(88)80091-3},
	number = {2},
	urldate = {2025-02-06},
	journal = {Powder Technology},
	author = {Erdész, K. and Szalay, A.},
	month = jun,
	year = {1988},
	pages = {87--96},
}

@techreport{sherrit_mechanism_2010,
	title = {Mechanism for {Particle} {Transport} and {Size} {Sorting} via {Low}-{Frequency} {Vibrations}},
	number = {NASA Tech Brief, Aug 2010},
	institution = {California Institute of Technology},
	author = {Sherrit, Stewart and Scott, James S. and Bar-Cohen, Yoseph and Badescu, Mircea and Bao, Xiaoqi},
	year = {2010},
}

@article{hui_vibrational_2024,
	title = {Vibrational manipulation of dry granular materials in lab-on-a-chip devices},
	volume = {24},
	issn = {1473-0197, 1473-0189},
	url = {https://xlink.rsc.org/?DOI=D3LC00722G},
	doi = {10.1039/D3LC00722G},
	number = {4},
	urldate = {2025-02-07},
	journal = {Lab on a Chip},
	author = {Hui, Timothy C. and Zhang, Xiaolin and Adiga, Dhruva and Miller, Gregory H. and Ristenpart, William D.},
	year = {2024},
	pages = {966--974},
}

@article{kim_forward_2020,
	title = {On the forward and backward motion of milli-bristlebots},
	volume = {127},
	issn = {00207462},
	url = {https://linkinghub.elsevier.com/retrieve/pii/S0020746220302134},
	doi = {10.1016/j.ijnonlinmec.2020.103551},
	urldate = {2025-02-07},
	journal = {International Journal of Non-Linear Mechanics},
	author = {Kim, D. and Hao, Z. and Mohazab, A.R. and Ansari, A.},
	month = dec,
	year = {2020},
	pages = {103551},
}

@article{kopitca_programmable_2021,
	title = {Programmable assembly of particles on a {Chladni} plate},
	volume = {7},
	issn = {2375-2548},
	url = {https://www.science.org/doi/10.1126/sciadv.abi7716},
	doi = {10.1126/sciadv.abi7716},
	number = {39},
	urldate = {2025-02-07},
	journal = {Science Advances},
	author = {Kopitca, Artur and Latifi, Kourosh and Zhou, Quan},
	month = sep,
	year = {2021},
	pages = {eabi7716},
}

@book{boussinesq_application_1885,
	address = {Paris},
	title = {Application des potentiels à l'étude de l'équilibre et des mouvements des solides elastiques (Application of potentials to the study of equilibrium and movements of elastic solids)},
	publisher = {Ganther-Villars},
	author = {Boussinesq, Joseph},
	year = {1885},
    note = {(in French)},
}

@book{viktorov_rayleigh_1966,
	address = {New York},
	title = {Rayleigh and Lamb Waves},
	publisher = {Springer New York},
	author = {Viktorov, I.A.},
	year = {1967},
    isbn = {978-1-4899-5683-5},
}

@article{aleshin_method_2015,
	title = {Method of memory diagrams for mechanical frictional contacts subject to arbitrary {2D} loading},
	volume = {60-61},
	issn = {00207683},
	url = {https://linkinghub.elsevier.com/retrieve/pii/S0020768315000566},
	doi = {10.1016/j.ijsolstr.2015.02.016},
	urldate = {2025-02-07},
	journal = {International Journal of Solids and Structures},
	author = {Aleshin, V. and Bou Matar, O. and Van Den Abeele, K.},
	month = may,
	year = {2015},
	pages = {84--95},
}

@article{mindlin_elastic_1953,
	title = {Elastic spheres in contact under varying oblique forces},
	volume = {20},
	journal = {Journal of Applied Mechanics},
	author = {Mindlin, R. D. and Deresiewicz, H.},
	year = {1953},
	pages = {327--344},
}

@article{jager_axi-symmetric_1995,
	title = {Axi-symmetric bodies of equal material in contact under torsion or shift},
	volume = {65},
	copyright = {http://www.springer.com/tdm},
	issn = {0939-1533, 1432-0681},
	url = {http://link.springer.com/10.1007/BF00835661},
	doi = {10.1007/BF00835661},
	number = {7},
	urldate = {2025-02-07},
	journal = {Archive of Applied Mechanics},
	author = {Jäger, J.},
	month = sep,
	year = {1995},
	pages = {478--487},
}

@book{landau_theory_1986,
	address = {Oxford},
	edition = {2. ed., rev. and enl, repr. 1986},
	series = {Course of theoretical physics},
	title = {Theory of elasticity},
	volume = {7},
	isbn = {978-0-08-006465-9},
	publisher = {Pergamon Press},
	author = {Landau, Lev Davidovič and Lifšic, Evgenij M.},
	year = {1986},
}

@book{popov_method_2015,
	address = {Berlin Heidelberg},
	title = {Method of {Dimensionality} {Reduction} in {Contact} {Mechanics} and {Friction}},
	publisher = {Springer-Verlag},
	author = {Popov, V. L. and Hess, M.},
	year = {2015},
}

@article{takizawa_manipulation_2021,
	title = {Manipulation of {Powder} with {Surface} {Acoustic} {Wave} {Actuator} to {Control} {Standing} and {Traveling} {Modes}},
	volume = {33},
	issn = {0914-4935, 2435-0869},
	url = {http://myukk.org/SM2017/article.php?ss=3489},
	doi = {10.18494/SAM.2021.3489},
	number = {12},
	urldate = {2025-02-07},
	journal = {Sensors and Materials},
	author = {Takizawa, Yukako and Fukuchi, Yusuke and Hamaguchi, Kazuya and Amaya, Satoshi and Utsumi, Yuichi and Takeo, Masahiro and Iimura, Kenji and Suzuki, Michitaka and Saiki, Tsunemasa},
	month = dec,
	year = {2021},
	pages = {4427},
}

@inproceedings{aleshin_contact_2025,
	title = {Contact {Forces} {Move} a {Particle} along an {Acoustically} {Excited} {Surface}: {Case} of {Permanent} {Contact}},
    address={Samarkand},
	volume = {3177},
	doi = {10.1063/5.0295665},
	booktitle = {{AIP} {Conf}. {Proc}.},
	author = {Aleshin, Vladislav V. and Terzi, Marina and Ghesquière, Jules},
	year = {2025},
	note = {Issue: 020001},
}

@article{hertz_uber_1881,
	title = {Über die berührung fester elastischer {Körper} (On the contact of solid elastic bodies)},
	volume = {92},
	journal = {Journal für die reine und angewandte Mathematik},
	author = {Hertz, H.},
	year = {1881},
	pages = {156--171},
    note = {(in German)},
}

@article{cattaneo_sul_1938,
	series = {6},
	title = {Sul {Contatto} die due corpi elastici: distributione locale degli sforzi. (On the contact of two elastic bodies: local stress distribution.)},
	volume = {27},
	journal = {Rendiconti dell'Aecademia Nazionale dei Lincei},
	author = {Cattaneo, C.},
	year = {1938},
	pages = {342--348,434--436,474--478},
    note = {(in Italian)},
}

@book{serway_physics_1996,
	address = {Philadelphia},
	edition = {4th ed},
	series = {Saunders golden sunburst series},
	title = {Physics for scientists {\&} engineers, with modern physics},
	isbn = {978-0-03-015654-0},
	publisher = {Saunders College Pub},
	author = {Serway, Raymond A.},
	year = {1996},
}

@book{hairer_solving_2000,
	address = {Berlin},
	edition = {2. rev. ed., 2. corr. print},
	series = {Springer series in computational mathematics},
	title = {Solving ordinary differential equations. 1: {Nonstiff} problems},
	isbn = {978-0-387-56670-2 978-3-540-56670-0},
	shorttitle = {Solving ordinary differential equations. 1},
	number = {8},
	publisher = {Springer},
	author = {Hairer, E. and Nørsett, S. P. and Wanner, G.},
	year = {2000},
}

@article{hongler_periodic_1989,
	title = {Periodic versus chaotic dynamics in vibratory feeders},
	volume = {62},
	journal = {Helvetica Physica Acta},
	author = {Hongler, M.-O. and Figour, J.},
	year = {1989},
	pages = {68--81},
}

@article{mracek_system_2005,
	title = {A system for powder transport based on piezoelectrically excited ultrasonic progressive waves},
	volume = {90},
	doi = {10.1016/j.matchemphys.2004.09.048},
	journal = {Materials Chemistry and Physics},
	author = {Mracek, M. and Wallaschek, J.},
	year = {2005},
	pages = {378--380},
}

@book{chladni_discoveries_1787,
	address = {Leipzig},
	title = {Discoveries in the {Theory} of {Sound}},
	publisher = {Weidmanns Erben \& Reich},
	author = {Chladni, E.},
	year = {1787},
}

@article{van_gerner_air-induced_2011,
	title = {Air-induced inverse {Chladni} patterns},
	volume = {689},
	url = {https://doi.org/10.1017/jfm.2011.411},
	journal = {Journal of Fluid Mechanics},
	author = {van Gerner, H. J. and van der Weele, K. and van der Hoef, M. A. and van der Meer, D.},
	year = {2011},
	pages = {203--220},
}

@article{zhou_controlling_2016,
	title = {Controlling the motion of multiple objects on a {Chladni} plate},
	volume = {7},
	url = {https://doi.org/10.1038/ncomms12764},
	number = {12764},
	journal = {Nature Communications},
	author = {Zhou, Q. and Sariola, V. and Latifi, K. and Liimatainen, V.},
	year = {2016},
}

@article{latifi_motion_2019,
	title = {Motion of {Heavy} {Particles} on a {Submerged} {Chladni} {Plate}},
	volume = {122},
	url = {https://doi.org/10.1103/PhysRevLett.122.184301},
	number = {184301},
	journal = {Physical Review Letters},
	author = {Latifi, K. and Wijaya, H. and Zhou, Q.},
	year = {2019},
}

@article{morita_simulation_1999,
	title = {Simulation of surface acoustic wave motor with spherical slider},
	volume = {46},
	doi = {10.1109/58.775659},
	number = {4},
	journal = {IEEE Trans Ultrason Ferroelectr Freq Control},
	author = {Morita, T. and Kurosawa, M. K. and Higuchi, T.},
	year = {1999},
	pages = {929--934},
}

@article{behera_design_2019,
	title = {Design and {Investigation} of a {Dual} {Friction}-{Drive}-{Based} {LiNbO3} {Piezoelectric} {Actuator} {Employing} a {Cylindrical} {Shaft} as {Slider}},
	volume = {19},
	doi = {10.1109/JSEN.2019.2938246},
	number = {24},
	journal = {IEEE Sensors Journal},
	author = {Behera, B.},
	year = {2019},
	pages = {11980--11987},
}

@techreport{bar-cohen_high-speed_2012,
	title = {High-{Speed} {Transport} of {Fluid} {Drops} and {Solid} {Particles} via {Surface} {Acoustic} {Waves}},
	number = {NASA Tech Briefs, December 2012},
	institution = {NASA’s Jet Propulsion Laboratory},
	author = {Bar-Cohen, Yoseph and Bao, Xiaoqi and Sherrit, Stewart and Badescu, Mircea and Lih, S.-S.},
	year = {2012},
}

@article{saiki_transporting_2021,
	title = {Transporting {Powder} with {Surface} {Acoustic} {Waves} {Propagating} on {Tilted} {Substrate}},
	volume = {33},
	url = {https://doi.org/10.18494/SAM.2021.3213},
	number = {12},
	journal = {Sensors and Materials},
	author = {Saiki, Tsunemasa and Takizawa, Yukako and Kaneyoshi, T. and Iimura, Kenji and Suzuki, Michitaka and Yamaguchi, A. and Utsumi, Yuichi},
	year = {2021},
	pages = {4409--4426},
}

@article{delrue_two_2018,
	title = {Two dimensional modeling of elastic wave propagation in solids containing cracks with rough surfaces and friction – {Part} {II}: {Numerical} implementation},
	volume = {82},
	issn = {0041624X},
	shorttitle = {Two dimensional modeling of elastic wave propagation in solids containing cracks with rough surfaces and friction – {Part} {II}},
	url = {https://linkinghub.elsevier.com/retrieve/pii/S0041624X17303700},
	doi = {10.1016/j.ultras.2017.07.003},
	urldate = {2025-05-10},
	journal = {Ultrasonics},
	author = {Delrue, Steven and Aleshin, Vladislav and Truyaert, Kevin and Bou Matar, Olivier and Van Den Abeele, Koen},
	month = jan,
	year = {2018},
	pages = {19--30},
}

@inproceedings{terzi_hopping_2025,
	address = {Paris},
	title = {Hopping motion of a particle along a substrate driven by a surface acoustic wave},
    booktitle = {Recueil des Résumés
et des Actes},
	url = {https://cfa2025.fr/pdfs/livret-resumes.pdf},
	author = {Terzi, Marina and Aleshin, Vladislav and Ghesquière, Jules},
	year = {2025},
	pages = {179--185},
}

@article{blekhman_effects_2016,
	title = {Effects produced by oscillations applied to nonlinear dynamic systems: a general approach and examples},
	volume = {83},
	issn = {0924-090X, 1573-269X},
	shorttitle = {Effects produced by oscillations applied to nonlinear dynamic systems},
	url = {http://link.springer.com/10.1007/s11071-015-2470-x},
	doi = {10.1007/s11071-015-2470-x},
	number = {4},
	urldate = {2025-09-09},
	journal = {Nonlinear Dynamics},
	author = {Blekhman, I. I. and Sorokin, V. S.},
	month = mar,
	year = {2016},
	pages = {2125--2141},
}

@book{popov_handbook_2019,
	address = {Berlin, Heidelberg},
	title = {Handbook of {Contact} {Mechanics}: {Exact} {Solutions} of {Axisymmetric} {Contact} {Problems}},
	copyright = {https://creativecommons.org/licenses/by/4.0},
	isbn = {978-3-662-58708-9 978-3-662-58709-6},
	shorttitle = {Handbook of {Contact} {Mechanics}},
	url = {http://link.springer.com/10.1007/978-3-662-58709-6},
	urldate = {2025-09-09},
	publisher = {Springer Berlin Heidelberg},
	author = {Popov, Valentin L. and Heß, Markus and Willert, Emanuel},
	year = {2019},
	doi = {10.1007/978-3-662-58709-6},
}

@article{willert_impact_2016,
	title = {Impact of an elastic sphere with an elastic half space with a constant coefficient of friction: {Numerical} analysis based on the method of dimensionality reduction},
	volume = {96},
	copyright = {http://onlinelibrary.wiley.com/termsAndConditions\#vor},
	issn = {0044-2267, 1521-4001},
	shorttitle = {Impact of an elastic sphere with an elastic half space with a constant coefficient of friction},
	url = {https://onlinelibrary.wiley.com/doi/10.1002/zamm.201400309},
	doi = {10.1002/zamm.201400309},
	number = {9},
	urldate = {2025-09-09},
	journal = {ZAMM - Journal of Applied Mathematics and Mechanics / Zeitschrift für Angewandte Mathematik und Mechanik},
	author = {Willert, E. and Popov, V. L.},
	month = sep,
	year = {2016},
	pages = {1089--1095},
}

@article{ragulskis_transport_2008,
	title = {Transport of particles by surface waves: a modification of the classical bouncer model},
	volume = {10},
	issn = {1367-2630},
	shorttitle = {Transport of particles by surface waves},
	url = {https://iopscience.iop.org/article/10.1088/1367-2630/10/8/083017},
	doi = {10.1088/1367-2630/10/8/083017},
	number = {8},
	urldate = {2025-09-09},
	journal = {New Journal of Physics},
	author = {Ragulskis, M and Sanjuán, M A F},
	month = aug,
	year = {2008},
	pages = {083017},
}

\end{document}